# Spintronic Thermal Management

Ken-ichi Uchida[1-3]* and Ryo Iguchi[1]

[1]*National Institute for Materials Science, Tsukuba 305-0047, Japan*
[2]*Institute for Materials Research, Tohoku University, Sendai 980-8577, Japan*
[3]*Center for Spintronics Research Network, Tohoku University, Sendai 980-8577, Japan*
*E-mail: UCHIDA.Kenichi@nims.go.jp

In spin caloritronics, a branch of spintronics, the spin degree of freedom is exploited for thermoelectric conversion and thermal transport. Since the inception of spin caloritronics, many experimental and theoretical studies have focused on clarifying the fundamental physics of the heat-to-spin and heat-to-charge current conversion phenomena in magnetic materials and magnetic hybrid structures, such as the spin Seebeck and anomalous Nernst effects. While research on these phenomena is progressing, there are also many spin-caloritronic phenomena that output heat currents. The observations of such phenomena have recently been accomplished through cutting-edge heat detection techniques. The recent developments in spin caloritronics have revealed that the generation, conversion, and transport of heat can be actively controlled by spins and/or magnetism. In this review article, we propose a new concept called *spintronic thermal management*. With proof-of-concept demonstrations, we introduce the basic principles, behaviors, measurement methods, and heat control functionalities of spin-caloritronic phenomena and discuss potential applications of spintronic thermal management.

## CONTENTS

1. **Introduction**

In recent years, electronic devices have decreased in size and their performance has improved. However, these benefits come at the cost of reduced reliability caused by high heat generation density, highlighting the importance of thermal management technologies.[1] Mainstream research topics related to thermal management technologies for electronic device applications include improving the performance of electronic cooling by thermoelectric effects,[2] creating materials with particularly high or low thermal conductivity,[1] improving interfacial heat transfer,[1,3] and artificially controlling the efficiency of thermal radiation.[4] Notably, the recent development of nanotechnology allows thermal conductivity to be reduced by nanostructuring and/or phonon band engineering.[5-7] These methods improve the efficiency of thermoelectric conversion because the figure of merit is inversely proportional to the thermal conductivity. To accelerate fundamental and application studies of thermal management technologies, new principles and materials should be introduced into thermoelectric conversion and thermal energy engineering.

Spintronics is a leading candidate for next-generation electronic technologies and has advanced in completely different directions from the above studies. Spintronics is a research field that aims to create new physical principles and applications by actively making use of the spin degree of freedom as an energy and information carrier.[8] The central topics in spintronics have long been the giant magnetoresistance (GMR) and tunnel magnetoresistance (TMR); since the discovery of these phenomena, spintronics has rapidly reached a practical level and GMR/TMR have been used as driving principles of highly sensitive magnetic sensors and nonvolatile memories.[9,10] As such studies progressed, researchers recognized the importance of the concept of a spin current, which is a flow of spin angular momentum, leading to the discovery of various transport phenomena driven or mediated by spins.[8] In this stream, thermal energy conversion technologies using spins and/or magnetism have been proposed, becoming a new trend in spintronics.[11,12]

The research field based on the combination of spintronics, thermoelectrics, and thermal energy engineering is called spin caloritronics, in which the interplay between spin, charge, and heat currents has been extensively investigated.[13-17] The rapid development of spin caloritronics was stimulated by the discovery of the spin Seebeck effect (SSE), which refers to the generation of a spin current from a heat current in magnetic materials.[18-21] The spin current generated by SSE can be converted into a charge current via spin-orbit interaction, enabling spin-current-driven thermoelectric generation.[11,12] Subsequently, the anomalous Nernst effect (ANE), one of the thermoelectric effects in magnetic materials, has attracted renewed attention because of the similar thermoelectric conversion symmetry to SSE[12,22,23] and its intriguing physical mechanism.[24-29] An important functionality realized by SSE and ANE is transverse thermoelectric generation, in which the input heat current and output charge current are perpendicular to each other.[12] This symmetry simplifies the structure of SSE and ANE devices compared to conventional Seebeck devices, in which the input heat current and output charge current are parallel to each other. Furthermore, in the case of SSE, thermoelectric generation from heat in an insulator is also possible.[11,12,19,21] Owing to these unique features, SSE and ANE have been actively investigated from the viewpoints of not only fundamental physics but also thermoelectric applications, although their thermopower and figure of merit are still small. Materials science studies to improve the thermoelectric conversion performance of SSE and ANE are in progress worldwide, with the aim of developing next-generation energy harvesting and heat sensing technologies.[11,12,30-34]

In magnetic materials and hybrid structures, various spin-driven and spin-mediated transport phenomena that output heat currents also exist.[15-17] Cutting-edge heat detection techniques simplify the measurements of such phenomena. For example, a recently developed thermoelectric imaging technique based on lock-in thermography (LIT)[35] enables the visualization of the temperature change induced by the spin Peltier effect[36] and the anomalous Ettingshausen effect,[37] which are the Onsager reciprocals of SSE and ANE, respectively. Furthermore, using the LIT technique, the direct observation of the anisotropic magneto-Peltier effect[38] was reported in 2018. As exemplified in these studies, thermoelectric conversion and thermal transport properties can be actively controlled by spins and/or magnetism.

Based on the recent developments in spin caloritronics, we propose a new concept called *spintronic thermal management*. This concept offers unique heat control functionalities, including local temperature modulation, active control of thermoelectric conversion, spintronic thermal switching, and unidirectional remote heating (Fig. 1). These functionalities arise from the intrinsic properties of magnetic materials, such as the spin-orbit interaction, spin-polarized electron transport, spin-wave or magnon transport, and quantum mechanical rectification, *i.e.*, nonreciprocity of spins. This review article aims to introduce the essence of spintronic thermal management and a new direction in spin caloritronics.

This review article is organized as follows. In Sect. 2, we summarize the definitions and basic behaviors of spin-caloritronic phenomena that form the basis of spintronic thermal management. In Sect. 3, we explain the heat detection principles and techniques used for investigating the spin-caloritronic phenomena that output heat currents. Experimental demonstrations of the heat control functionalities made possible by spin caloritronics are reviewed in Sect. 4. Research activities on materials science and device engineering are shown in Sects. 5 and 6, respectively, which are necessary for applications of spintronic thermal management. Section 7 presents our conclusion and prospects. We anticipate that the introduction of the concept of spintronic thermal management will invigorate basic science and application studies on spin caloritronics.

## 2. Heat Transport Phenomena in Spin Caloritronics

In spin caloritronics, various transport phenomena based on the interplay among spin, charge, and heat currents have been discovered. The terminologies of spin-caloritronic phenomena are often confusing because of their similarities.[15-17] Thus, we first classify spin-caloritronic phenomena into three categories: (1) magneto-thermoelectric effects, (2) thermomagnetic effects, and (3) thermospin effects (Fig. 2). The magneto-thermoelectric effect refers to the conversion between charge and heat currents in a conductor under a magnetic field or in a magnetic material with spontaneous magnetization, where the thermoelectric conversion properties depend on the magnetic field or magnetization. In this article, the thermomagnetic effect represents the phenomenon in which heat transport, *i.e.*, thermal conductivity, depends on a magnetic field or magnetization, although some studies use that term to refer to the magneto-thermoelectric effect. The thermospin effect refers to the conversion between spin and heat currents, which is conceptually different from the magneto-thermoelectric effects despite the similar symmetries and functionalities. Notably, since the discovery of SSE, a wide variety of thermospin effects have been discovered in magnetic materials and their junction structures, because spin currents are carried not only by conduction electrons but also by the collective dynamics of local magnetic moments, *i.e.*, spin waves or magnons (Fig. 2).[8] In SSE and the spin Peltier effect, the thermally generated spin current is carried by incoherent spin waves or thermal/subthermal magnons governed by the exchange interaction. In contrast, spin waves governed by the dipole-dipole interaction can also carry energy and eventually emit heat. The dipole-dipole interaction is responsible for the dynamics of spin waves at low energies, accessible by using microwaves. Such spin waves are referred to as magnetostatic spin waves.

In the following subsections, we introduce the basic behaviors of the spin-calorinic phenomena. In addition to the aforementioned categorization, electron transport phenomena in magnetic materials are categorized into longitudinal and transverse effects, in which the output current is generated parallel and perpendicular to the input current, respectively.[17)] In Sect. 2.1 (2.2), we explain the symmetries of the longitudinal and transverse magneto-thermoelectric (thermomagnetic) effects. Here, we omit detailed explanations about magnetic-field-dependent magneto-thermoelectric and thermomagnetic effects and focus on magnetization-dependent effects because of their nonvolatility and applicability to spintronic devices. In Sect. 2.3, we present the fundamentals of the thermospin effects.

*2.1 Magneto-thermoelectric effects*

The longitudinal and transverse magneto-thermoelectric effects that can be applied to spintronic thermal management are the anisotropic magneto-Peltier effect (AMPE) and the anomalous Ettingshausen effect (AEE), respectively [Fig. 3(a)]. These phenomena appear in magnetic materials even in the absence of an external magnetic field, owing to the action of the spin-orbit interaction on spin-polarized conduction electrons. Although AMPE and AEE have long been known, they remained unstudied because of the difficulty involved in measuring them. This was solved by the LIT technique detailed in Sect. 3.4; now, AMPE and AEE can be investigated easily in simple sample structures. Recent studies have clarified that AMPE and AEE exhibit various unique heat control functionalities that cannot be achieved with the conventional Peltier effect alone.

*2.1.1 Anisotropic magneto-Peltier effect*

AMPE is a phenomenon in which the Peltier coefficient $\Pi$ of a magnetic material changes depending on the relative angle $\theta$ between the magnetization $\mathbf{M}$ and the applied charge current $\mathbf{J}_c$. Thus, if the material exhibits AMPE, the heat current generated along the charge current in the $\mathbf{M} \perp \mathbf{J}_c$ configuration is different from that in the $\mathbf{M} \parallel \mathbf{J}_c$ configuration [see the top left quadrant in Fig. 3(a)]; the discontinuity of the heat currents becomes heat sources in the material. The $\theta$ dependence of $\Pi$ in an isotropic magnetic material is expressed as

$$\Pi(\theta) = \Pi_\perp + (\Pi_\parallel - \Pi_\perp)\cos^2\theta, \tag{1}$$

where $\Pi_\perp$ ($\Pi_\parallel$) is the Peltier coefficient for the $\mathbf{M} \perp \mathbf{J}_c$ ($\mathbf{M} \parallel \mathbf{J}_c$) configuration. Therefore, when the magnetization distribution is nonuniform and/or the path of the charge current is bent, AMPE can induce heat release or absorption at the boundaries between the regions with different $\theta$ values, enabling thermoelectric heating or cooling in single materials without junction structures. In 2016, AMPE was indirectly detected in a nonlocal spin valve structure comprising $Ni_{80}Fe_{20}$/Al junctions by Das *et al*.[39)] In 2018, we directly observed the AMPE-induced temperature change in a Ni slab without junctions using the LIT technique.[38)] Direct observation of AMPE was accomplished by forming a U-shaped ferromagnetic metal. Even when the U-shaped structure is uniformly magnetized, a nonuniform $\theta$ distribution can occur owing to the different $\mathbf{J}_c$ directions; in the configuration depicted in Fig. 3(b), the region between the corners (the leg part) possesses $\Pi_\parallel$ ($\Pi_\perp$). Thus, both corners of the U-shaped structure become the boundaries between the regions with $\Pi_\perp$ and $\Pi_\parallel$, producing the AMPE-induced heat release and absorption when $\Pi_\parallel - \Pi_\perp$ ($\equiv \Delta\Pi$) is finite.

AMPE may be utilized in constructing versatile thermoelectric cooling devices if a large $\Delta\Pi$ is obtained. The most important feature of AMPE is the generation of heat release and absorption in single materials without junctions, which makes thermoelectric devices much simpler and more cost-effective. Furthermore, the thermoelectric conversion characteristics of AMPE devices can be reconfigured by changing the $\mathbf{M}$ distribution and/or shape of the magnetic materials. This feature is in stark contrast to conventional Peltier devices, the thermoelectric conversion characteristics of which cannot be modified once they are constructed. We note that, in AMPE devices with appropriately designed magnetization distributions, U-shape or bent shapes are not necessarily required, and temperature modulation can be generated at arbitrary positions. These AMPE features are useful for pinpoint temperature control, allowing the direct electronic cooling of a small region at which conventional Peltier devices comprising multiple junctions cannot be integrated. Nevertheless, the figure of merit for AMPE is still much smaller than that for the Peltier effect. Therefore, investigations are being pursued in materials science and device engineering to take advantage of the AMPE features.

AMPE occurs when the spin polarization direction of conduction electrons, *i.e.*, the magnetization direction, and the charge current direction are coupled via the spin-orbit interaction in magnetic materials. The Peltier coefficient in metals is often discussed based on the Mott formula:

$$\Pi = -\frac{\pi^2 k_B{}^2 T^2}{3e} \frac{1}{\sigma(\varepsilon_F)} \left[\frac{\partial \sigma(\varepsilon)}{\partial \varepsilon}\right]_{\varepsilon=\varepsilon_F}, \qquad (2)$$

where $k_B$ is the Boltzmann constant, $T$ is the absolute temperature, $e$ ($> 0$) is the elementary charge, $\sigma(\varepsilon)$ is the energy ($\varepsilon$)-dependent electrical conductivity, and $\varepsilon_F$ is the Fermi energy. Owing to the spin-orbit interaction, $\Pi$ and $\sigma$ depend on $\theta$. When the anisotropies of the Peltier coefficient and electrical conductivity are small, $\Delta\Pi$ is approximately expressed as

$$\frac{\Delta\Pi}{\Pi} \approx \left(\frac{\sigma_\perp - \sigma_\parallel}{\sigma_\perp}\right) - \left\{\frac{[\partial \sigma_\perp(\varepsilon)/\partial \varepsilon]_{\varepsilon=\varepsilon_F} - [\partial \sigma_\parallel(\varepsilon)/\partial \varepsilon]_{\varepsilon=\varepsilon_F}}{[\partial \sigma_\perp(\varepsilon)/\partial \varepsilon]_{\varepsilon=\varepsilon_F}}\right\}, \qquad (3)$$

where $\sigma_\perp$ ($\sigma_\parallel$) is the electrical conductivity for the $\mathbf{M} \perp \mathbf{J}_c$ ($\mathbf{M} \parallel \mathbf{J}_c$) configuration.[40] The first term on the right side of Eq. (3) is the contribution from the anisotropic magnetoresistance.[41] Therefore, if a magnetic material exhibits the anisotropic magnetoresistance, it should also exhibit AMPE, unless the first term cancels out the second term. The contribution of the second term often plays an essential role in AMPE, and materials are needed in which the energy derivative of $\sigma(\varepsilon)$, or the density of states, at the Fermi energy is anisotropic due to the spin-orbit interaction. In Ni$_{95}$Pt$_5$, which exhibits the largest AMPE reported so far, the anisotropy of the Peltier coefficient due to the second term reaches 13%, whereas that due to the first term is only 1%.[40] However, the material dependence of AMPE is not simple. For example, although the doping of Ni with a small amount of Pt increases $\Delta\Pi$, excess Pt doping decreases $\Delta\Pi$.[40,42] In contrast to the Ni-based alloys, almost no AMPE signals were observed for Fe and FePt.[40] Each term in Eq. (3) can be estimated experimentally, but no clear material design guideline to enhance AMPE is available. Thus, finding and developing magnetic materials with large $\Delta\Pi$ are significant challenges. In 2019, theoretical work to address this issue was reported by Masuda *et al*.[43] Through first-principles calculations, they reproduced the experimental results of AMPE for Ni and Fe and proposed several promising ordered alloys with giant $\Delta\Pi$. Experimental studies to create magnetic materials with large AMPE are also being conducted.

*2.1.2 Anomalous Ettingshausen effect*

AEE is the transverse magneto-thermoelectric effect that outputs a heat current in a magnetic material. In AEE, when $\mathbf{J}_c$ is applied to a magnetic material in the direction perpendicular to $\mathbf{M}$, a heat current is generated in the direction of the vector product of $\mathbf{J}_c$ and $\mathbf{M}$ [see the top right quadrant in Fig. 3(a)]. Thus, the symmetry of AEE is expressed as

$$\mathbf{j}_{q,\text{AEE}} = \Pi_{\text{AEE}} \left(\mathbf{j}_c \times \frac{\mathbf{M}}{|\mathbf{M}|}\right), \qquad (4)$$

where $\Pi_{\text{AEE}}$ is the anomalous Ettingshausen coefficient, $\mathbf{j}_{q,\text{AEE}}$ is the heat current density generated by AEE, and $\mathbf{j}_c$ is the charge current density. $\Pi_{\text{AEE}}$ represents the thermoelectric performance of AEE and satisfies the relation $\Pi_{\text{AEE}} = S_{\text{ANE}} T$, where $S_{\text{ANE}}$ is the anomalous Nernst coefficient. This is the Onsager reciprocal relation between AEE and ANE, meaning that materials with large ANE also have large AEE. Equation (4) represents the heat control functionalities unique to AEE: (i) the input charge current and output heat current are perpendicular to each other and (ii) the direction of the heat current can be controlled by tuning the $\mathbf{M}$ direction or distribution. Feature (i) enables the simplification of the thermoelectric device structure (see Sect. 6). Using feature (ii), the active control of heat currents can be achieved by controlling the magnetism by external forces or fields. The demonstrations of the AEE-specific functionalities are presented in Sect. 4.

The research history of AEE is summarized as follows. The Ettingshausen effect in typical bulk ferromagnetic metals was reported approximately 100 years ago.[44-46] However, in the pioneering studies, the magnetization-dependent component was not separated from the magnetic-field-dependent component, and the temperature distribution induced by the Ettingshausen effect was not confirmed. Since then, no studies on AEE were reported until the LIT technique started to be used in spin caloritronics. As discussed later, the LIT technique enables the observation of AEE in both bulk and thin film systems and the visualization of the AEE-induced temperature modulation. The first observation of AEE in thin films was reported by Seki *et al*. in 2018.[37] They clearly demonstrated the difference in the thermoelectric conversion symmetries between AEE and the spin Peltier effect (see Sect. 4.1). The U-shaped structure is also useful in the LIT measurements of AEE because the symmetry described in Eq. (4) can be verified simultaneously. Although many magnetic materials exhibit both AEE and AMPE, their contributions can be separated by measuring the magnetic field dependence of the current-induced temperature change because AEE (AMPE) signals exhibit odd (even) dependence on the **M** direction [see Eqs. (1) and (4)].[38,47] The thermoelectric performance of AEE, *i.e.*, $\Pi_{AEE}$ and figure of merit, can now be evaluated quantitatively with simple sample preparation, measurement, and analysis procedures.[40,48,49]

In recent years, physics and materials science studies on AEE/ANE have progressed rapidly with the advancement of spin caloritronics and topological materials science. In 2018, Sakai *et al*. reported the observation of large ANE in the magnetic Heusler compound $Co_2MnGa$ and found that large transverse thermopower is attributed to the topology of the electronic structure called the Weyl points.[25] At present, $S_{ANE}$ = 6–8 µVK$^{-1}$ observed for this material is the record-high value at temperatures above room temperature, which is more than an order of magnitude larger than $S_{ANE}$ for typical ferromagnetic metals, such as Fe and Ni.[34] Subsequently, the observation of large ANE in $Co_2MnGa$ and other Heusler compounds has been reported by several research groups, and studies on ANE using magnetic topological materials have become a global trend.[26-29] Because of the reciprocity between ANE and AEE, such Heusler compounds are also promising for thermal management technologies based on AEE. In parallel with the ANE studies, we investigated AEE in various ferromagnetic binary alloys and rare-earth magnets.[40,48,49] The LIT measurements reveal that a $SmCo_5$-type magnet, one of the practical permanent magnets, exhibits large AEE: $S_{ANE}$ was estimated to be 3–6 µVK$^{-1}$ based on the Onsager reciprocal relation.[48,49] Because permanent magnets are mass-produced and AEE/ANE in magnets with large remanent magnetization and coercivity work even in the absence of external magnetic fields, adding thermoelectric conversion functionalities to permanent magnets results in advantages to the applications of spin caloritronics. Importantly, the thermoelectric conversion performance of AEE/ANE has no correlation with the saturation magnetization in many magnetic materials since the intrinsic mechanism due to electronic structures is important in these phenomena.[34] Therefore, to develop magnetic materials with large AEE/ANE, unconventional knowledge and techniques are needed. In contrast, the lack of correlation between ANE/AEE and the saturation magnetization is convenient for thermoelectric applications because magnetic materials with large $S_{ANE}$ and small saturation magnetization are preferable for constructing ANE/AEE-based thermoelectric modules.[34]

*2.1.3 Peltier-driven transverse thermoelectric cooling*

In 2021, Zhou *et al*. proposed and demonstrated transverse thermoelectric generation different from ANE, which appears in a hybrid structure comprising a thermoelectric semiconductor and a magnetic metal.[50] Because this effect originates from the combination of the Seebeck effect in the thermoelectric semiconductor and the anomalous Hall effect in the magnetic metal, they called it Seebeck-driven transverse thermoelectric generation. The phenomenological model calculation shows that, by optimizing the combination of the thermoelectric semiconductor and magnetic metal as well as their dimensions, the Seebeck-driven transverse thermopower can reach the order of 100 µVK$^{-1}$. In fact, Zhou *et al*. show that the $Co_2MnGa$/n-type Si ($Co_2MnGa$/p-type Si) hybrid material exhibits a transverse thermopower of 82.3 µVK$^{-1}$ (−41.0 µVK$^{-1}$), which is one order of magnitude larger than the record-high $S_{ANE}$.[50] Although the construction of such hybrid materials partially hinders the merits of the transverse thermoelectric conversion, the Seebeck-driven transverse thermoelectric generation may be a breakthrough approach to spin caloritronics applications owing to its large thermopower.

Subsequently, Yamamoto *et al.* formulated the reciprocal process of the Seebeck-driven transverse thermoelectric generation in a closed circuit comprising a thermoelectric semiconductor and a magnetic metal.[51] This process should be called Peltier-driven transverse thermoelectric cooling; when a charge current is applied to the magnetic metal of the closed circuit, the anomalous Hall effect induces a transverse charge current in the thermoelectric semiconductor. The transverse charge current then generates heat release and absorption at the junctions of the thermoelectric semiconductor and magnetic metal owing to the difference in the Peltier coefficients between them. Yamamoto *et al.* showed that the coefficient of performance for this transverse thermoelectric cooling can be larger than that for AEE by parameter optimization in a manner similar to the Seebeck-driven transverse thermoelectric generation. As discussed here, not only development of magnetic materials with large AEE/ANE and the anomalous Hall effect but also optimization of thermoelectric/magnetic hybrid materials is essential for spintronic thermal management.

*2.2 Thermomagnetic effects*

The longitudinal and transverse thermomagnetic effects are the magneto-thermal resistance (MTR) and thermal Hall effect, respectively [Fig. 3(a)]. These phenomena have been known for a long time but are still far from being applied. Nevertheless, MTR and the thermal Hall effect have potential use in thermal switching devices and heat current circulators, respectively.

*2.2.1 Magneto-thermal resistance*

MTR refers to the magnetic field or magnetization dependence of the thermal conductivity, which is often called the Maggi-Righi-Leduc effect.[52] Thus, MTR in a plain magnetic material is regarded as the thermal analogue of the anisotropic magnetoresistance; the thermal conductivity in the $\mathbf{M} \perp \mathbf{J}_q$ configuration is different from that in the $\mathbf{M} \parallel \mathbf{J}_q$ configuration, where $\mathbf{J}_q$ is the heat current [see the bottom left quadrant in Fig. 3(a)].[53] However, the $\mathbf{M}$-dependent anisotropy of the thermal conductivity is usually small in ferromagnetic metals, in a manner similar to that of the anisotropic magnetoresistance.

The importance of MTR as a principle of spintronic thermal management emerges in magnetic multilayer films comprising alternately stacked ferromagnetic metals and nonmagnetic materials. Such multilayer films are the core of spintronic devices because they exhibit GMR (TMR) when the nonmagnetic layer is a metal (an insulating tunnel barrier).[9,10] In magnetic multilayer films, the electrical conductivity is larger when the $\mathbf{M}$ directions of the adjacent ferromagnetic layers are aligned parallel than when they are antiparallel because of GMR or TMR, where the $\mathbf{M}$ configuration can be switched by an external magnetic field or spin injection. It is known that the thermal conductivity also depends on the $\mathbf{M}$ configuration [Fig. 4(a)]: this is MTR in magnetic multilayer films.[54-57] Recent experiments showed that the thermal conductivity switching ratio $(\kappa_P - \kappa_{AP})/\kappa_{AP}$ in magnetic metal multilayers can exceed 100% even above room temperature, where $\kappa_P$ ($\kappa_{AP}$) is the thermal conductivity for the parallel (antiparallel) magnetization configuration.[57] Surprisingly, the thermal conductivity switching ratio can be larger than the magnetoresistance ratio, *i.e.*, the electrical conductivity switching ratio. MTR was observed not only in current-perpendicular-to-plane GMR devices, depicted in Fig. 4(a), but also in current-in-plane GMR devices. Because MTR in magnetic multilayers is applicable directly to the thermal management of spintronic devices, further enhancement of the thermal conductivity switching ratio and elucidation of its microscopic mechanism are important. The details of the recent experimental results on MTR in current-perpendicular-to-plane GMR systems are presented in Sect. 4.3.

*2.2.2 Thermal Hall effect*

The thermal Hall effect is a thermal analogue of the Hall effect, which is often called the Righi-

Leduc effect.[52] Following the culture of the Hall effects, the magnetic-field-dependent and magnetization-dependent thermal Hall effects can be called the ordinary and anomalous thermal Hall effects, respectively [see the bottom right quadrant in Fig. 3(a)]. When a material exhibits the thermal Hall effect, a heat current is bent in the direction perpendicular to an external magnetic field or magnetization, where the direction of the transverse heat current is reversed by reversing the magnetic field or magnetization. This feature can be regarded as a heat current circulator, which provides a principle of thermal switching (Fig. 1). Importantly, the thermal Hall effect appears not only in conductors but also in insulators, where the carriers of heat currents are phonons and/or magnons.[58-60]

*2.3 Thermospin effects*

In spin caloritronics, a variety of thermospin effects have been investigated. The diversity of the thermospin effects comes from the fact that a spin current is transported by various spin carriers, such as conduction electrons, magnons,[19,21] and spinons.[61] Since 2008, many experimental and theoretical studies have focused on the spin-dependent Seebeck effect and SSE, which are the conversion of heat currents into spin currents carried by conduction electrons and magnons, respectively. The Onsager reciprocals of these phenomena are the spin-dependent Peltier effect and the spin Peltier effect.

*2.3.1 Spin-dependent Peltier effect*

The spin-dependent Peltier effect refers to the conversion of a conduction electron spin current into a heat current.[15-17] The origin of the spin-dependent Peltier effect is the spin dependence of the Peltier coefficient in ferromagnetic metals. Here, let us recall that the density of states and its energy derivative at the Fermi energy for up-spin electrons are different from those for down-spin electrons in ferromagnetic metals. Because the Peltier coefficient is determined mainly by these parameters in semi-classical theory [Eq. (2)], it is natural that the coefficient for up-spin electrons is different from that for down-spin electrons. Thus, if a conduction electron spin current is injected into a ferromagnetic metal, up- and down-spin electrons carry different amounts of heat in opposite directions, leading to the generation of a net heat current.

The first direct observation of the spin-dependent Peltier effect was reported by Flipse *et al.* in 2012.[62] They measured the charge-current-induced temperature change in a nanopillar spin-valve device consisting of two ferromagnetic metal layers separated by a nonmagnetic metal, *i.e.*, a current-perpendicular-to-plane GMR device, using an on-chip thermocouple sensor. In this experiment, the current-induced temperature change was observed to be dependent on the magnetization configuration. This behavior results from the distributions of the spin-dependent electrochemical potential and temperature in the current-perpendicular-to-plane GMR device in the parallel magnetization configuration being different from those in the antiparallel magnetization configuration due to the spin-dependence of the Peltier coefficient. A similar behavior was also observed in a magnetic tunnel junction, *i.e.*, a TMR device, in the current-perpendicular-to-plane configuration.[63] These experiments confirm the existence of the spin-dependent Peltier effect and its physics are now well established. However, because the measurements using the GMR and TMR devices were performed in the presence of a charge current bias, the observed effects are usually called the magneto-Peltier effect [Fig. 4(b)]. In principle, the spin-dependent Peltier effect can also be driven by a pure spin current induced by the spin Hall effect.[64,65] It is worth noting that the direct observation of the magneto-Peltier effect in a current-in-plane GMR device was reported in 2019,[66] although its origin cannot be discussed in terms of the spin-dependent Peltier effect (recall the difference in the mechanism between current-in-plane and current-perpendicular-to-plane GMRs).

*2.3.2 Spin Peltier effect*

The spin Peltier effect (SPE) refers to the conversion of a magnon spin current into a heat current.[15-17] Despite the similar names and functionalities, the mechanism of SPE is completely different from that of the spin-dependent Peltier effect (Fig. 2). The significant difference between these phenomena is that the SPE appears not only in metals and semiconductors but also in insulators because of the magnon origin, while the spin-dependent Peltier effect appears only in conductors. A typical system used for measuring SPE is a junction structure consisting of a ferrimagnetic insulator (*e.g.*, yttrium iron garnet: $Y_3Fe_5O_{12}$) and a paramagnetic metal with large spin-orbit interaction (*e.g.*, Pt). When $\mathbf{J}_c$ is applied to the paramagnetic metal layer, a conduction electron spin current is generated in the direction of the film thickness by the spin Hall effect, and the spin accumulation occurs near the metal/insulator interface. The spin polarization vector **σ** of the spin accumulation is determined by the symmetry of the spin Hall effect:

$$\mathbf{J}_s \propto \mathbf{\sigma} \times \mathbf{J}_c, \tag{5}$$

where $\mathbf{J}_s$ is the spatial direction of the spin current. When **σ** in the paramagnetic metal layer is parallel or antiparallel to **M** of the ferrimagnetic insulator, this spin accumulation is converted into a magnon spin current in the ferromagnetic insulator via the interfacial exchange interaction, *i.e.*, spin-mixing conductance [Figs. 2 and 5(a)]. The magnon spin current then induces $\mathbf{J}_q$ in the vicinity of the interface due to SPE. The SPE-induced $\mathbf{J}_q$ obeys

$$\mathbf{J}_q \propto (\mathbf{\sigma} \cdot \mathbf{M})\,\mathbf{n}, \tag{6}$$

where **n** is the normal vector of the junction interface. Equations (5) and (6) show that the conversion symmetry between $\mathbf{J}_c$ and $\mathbf{J}_q$ in SPE is similar to that in AEE when **σ** and **M** are parallel (Fig. 5). In contrast, no SPE-induced temperature change appears when **σ** and **M** are orthogonal (see Sect. 4.1).[37]

The measurement of SPE requires highly sensitive and accurate temperature detection because the spin-to-heat current conversion appears within the scale of the magnon propagation length (~ 1 μm for $Y_3Fe_5O_{12}$). The first experimental observation of SPE was also reported by Flipse *et al.* in 2014, in which the temperature change in $Pt/Y_3Fe_5O_{12}$ junction systems was detected with micro thermocouple sensors patterned by electron beam lithography.[67] However, since its discovery, no experimental studies on SPE were reported for several years because of the measurement difficulty. This situation was changed by the LIT technique; in 2016, Daimon *et al.* demonstrated thermal imaging of SPE-induced temperature modulation using LIT and found its unique spatial distribution (see Sect. 4.1).[36] The LIT method enables the measurement of SPE with high temperature and spatial resolutions in simple sample structures that can be prepared without using micro/nanofabrication techniques. Based on the knowledge obtained from the LIT measurements, SPE was subsequently observed using a simple thermocouple wire[68] as well as a lock-in thermoreflectance (LITR) method.[69,70] These measurement techniques are explained in detail in Sect. 3.

*2.3.3 Transverse thermospin effect*

In the spin-dependent Peltier effect and SPE, the flow directions of spin and heat currents are parallel. Thus, these phenomena can be classified as longitudinal thermospin effects. Recent studies on spin caloritronics have revealed that transverse thermospin effects also exist. In 2017, several groups reported the observation of the spin Nernst effect, which is the thermal analogue of the spin Hall effect.[71-73] These experiments suggest the existence of its reciprocal: the spin Ettingshausen effect, where a transverse heat current is generated by a conduction electron spin current injected into a nonmagnetic conductor via the spin-orbit interaction. However, the temperature change induced by the spin Ettingshausen effect is yet to be observed directly. With further developments in thermal measurement and analysis techniques, the physics and functionalities realized by the transverse thermospin effects will be clarified.

*2.3.4 Spin-wave heat conveyer effect*

As discussed in Sect. 2.3.2, spin waves (or magnons) can carry heat and thus induce a heat current. To discuss the spin-wave-induced heat current, it is convenient to consider the spin-wave spin current:

$$\mathbf{j}_s = \hbar \sum_\mathbf{k} n(\mathbf{k})\mathbf{v}(\mathbf{k}), \tag{7}$$

where **k** denotes the wave number or momentum, $n$ is the population, and **v** is the group velocity of spin waves. This means that the imbalance between spin waves going to the right (+**k**) and left (–**k**) results in finite $\mathbf{j}_s$. Such an imbalance in the exchange spin waves can be induced by spin current injection and temperature gradients, as in the cases of SPE and SSE, respectively. In addition, an imbalance in the magnetostatic spin waves can be induced by microwaves under spin-wave resonance conditions.

One of the unique features of magnetostatic spin waves is the nonreciprocity appearing in magnetostatic surface spin waves, which are often called the Damon-Eshbach modes. These modes are localized at the surfaces of magnetic materials and their propagation direction is determined by the surface normal $\mathbf{n}$ and the $\mathbf{M}$ direction; the allowed $\mathbf{k}$ satisfies the relation $\mathbf{k} \parallel \mathbf{M} \times \mathbf{n}$. Consequently, on a specified surface, the magnetostatic surface spin waves flow in one direction and their backscattering seldom occurs, resulting in the generation of finite $\mathbf{j}_s$. Interestingly, the propagation direction of the magnetostatic surface spin waves can be controlled by the $\mathbf{M}$ direction. By exciting the magnetostatic surface spin waves on one surface by a localized microwave field via antennae, the spin-wave spin current concomitant with a heat current is generated, indicating that the direction of the heat current can also be controlled by the $\mathbf{M}$ direction. This phenomenon is called the unidirectional spin-wave heat conveyer effect, discovered by An *et al*. in 2013 (Fig. 6).[74] The heat control functionalities realized by this effect are shown in Sect. 4.4.

## 3. Measurement Techniques

Heat detection techniques play a major role in elucidating the physics of the spin-caloritronic phenomena, demonstrating their heat control functionalities, and discovering useful materials. There are two important techniques for investigating the spin-caloritronic phenomena that output heat currents. One is lock-in temperature detection, which is indispensable for detecting the temperature modulation signals induced by the spin-caloritronic phenomena with high temperature resolution and for distinguishing their contributions from parasitic effects. The other is a thermal imaging technique, which is useful for demonstrating the symmetries and functionalities of the spin-caloritronic phenomena. A technique that combines both lock-in and imaging temperature measurements is the LIT method, which was introduced into the studies of spin caloritronics in 2016.[36,75] As a complementary technique to LIT, the LITR method has recently been used to measure the magneto-thermoelectric and thermospin effects.[69,70] In the LIT and LITR methods, it is relatively easy to obtain highly sensitive temperature measurements with a temperature resolution of < 1 mK by selectively extracting the temperature modulation that follows the periodic excitation using Fourier analysis. As shown in Fig. 7, there are various options for the lock-in sources, such as charge and spin currents,[36-38,40,47-49] magnetic fields,[76,77] and light,[78] and various physical phenomena that respond to the input energies can be measured precisely. In the following subsections, we summarize the measurement principles and features of the temperature measurement techniques used in spin caloritronics and show some examples of the experimental results.

*3.1 Thermocouple*

A thermocouple is a popular contact-type temperature sensor that comprises a junction of two different metallic wires with different Seebeck coefficients. Attaching the junction to a target area on a sample induces an electric voltage between the ends of the two wires, with a magnitude proportional to the temperature difference between the junction and the ends of the thermocouple. The temperature of the target area can be estimated when the junction and the target area are thermalized and the ends of the thermocouple are thermally anchored to a heat bath. This simple method possesses sufficient sensitivity for detecting the spin-caloritronic phenomena. In fact, the first observation of SPE was performed using an on-chip thermocouple sensor prepared by a microfabrication technique.[67] The temperature sensitivity can be improved by constructing a thermopile structure consisting of an array of many thermocouples, where the output voltage is proportional to the number of individual thermocouples.

Figure 8 shows an example of the thermocouple-based detection of SPE for a Pt/Y$_3$Fe$_5$O$_{12}$ junction system, where a thermocouple was attached to the top of the Pt strip.[68] Here, a very thin thermocouple wire with a diameter of 13 μm was used to reduce the temperature stabilization time and was thermally connected to the substrate to maintain a stable reference temperature [Fig. 8(a)]. By detecting the difference between the voltage signals for positive and negative charge currents applied to the Pt strip, the temperature change induced by SPE $\Delta T_{SPE}$ was clearly observed [note that the direction of $\sigma$ and sign of $\Delta T_{SPE}$ are reversed by reversing $\mathbf{J}_c$, as described in Eqs. (5) and (6)]. The sign of the temperature change was reversed by reversing the magnetic field in response to the magnetization reversal of $Y_3Fe_5O_{12}$, indicating that the SPE contribution was successfully distinguished from other thermoelectric signals [Fig. 8(b)]. The thermocouple-based method is applicable to measurements of wide-range temperature and/or magnetic field dependences if the output of the thermocouple is appropriately calibrated. The suppression of $\Delta T_{SPE}$ observed in the high field range in Fig. 8(b) is attributed to the suppression of subthermal magnon excitations by the Zeeman gap.[79,80] Such high-magnetic-field responses cannot be measured by the LIT method because an infrared camera does not work well under high magnetic fields.

Although thermocouple-based measurements are very simple, the coupling between the thermocouple and the sample is often problematic. When the junction and sample surface are weakly connected, it takes a longer time for the temperature distribution to reach a steady state. Furthermore, the resultant thermal responses can be reduced when heat leakage through the thermocouples is significant. Thus, the experimental setup requires special attention in most cases.

*3.2 Thermography*

In the thermography method, the temperature distribution is measured via the infrared intensity emitted from the surface of materials. According to the Stefan-Boltzmann law, the total amount of radiant energy is proportional to the fourth power of thermodynamic temperature. At room temperature, low-energy infrared radiation predominates over the spectrum; thus, its intensity distribution can be imaged using an array of infrared sensors, such as InSb and HgCdTe photodetectors and microbolometers. By converting the radiation intensity into temperature information, noncontact thermal imaging can be performed, making measurements stable and reliable. Furthermore, thermography is useful for determining heat source positions and symmetries of target phenomena.

The quantitativeness of the thermography method depends on the infrared emissivity of target materials in the wavelength range to be detected. Because there is no perfect black body, it is necessary to obtain a calibration curve, *i.e.*, a relation between infrared intensity and temperature. When the sample is transparent in the detection wavelength range, calibration does not work. To exclude this possibility, the sample surface is usually coated with materials with high infrared emissivity, such as carbon and oxide-based insulating black ink, with a typical emissivity of > 0.95.

*3.3 Thermoreflectance*

Thermoreflectance refers to the change in optical reflectivity in response to temperature change. By irradiating a target area of a sample with light and detecting the reflection intensity, the temperature can be estimated in a noncontact manner. Typically, the temperature derivative of the reflectivity is small, and a photodetector and voltmeter with a high dynamic range are required. Thermal imaging measurements based on thermoreflectance are possible using cameras for visible light, although the temperature resolution is not high.[81]

Unlike infrared thermography, this method requires a light source and offers several measurement variations. For example, thermoreflectance thermometry can be combined with an optical pump-and-probe technique to study transition responses of thermal conduction in a short time scale ranging from picoseconds to nanoseconds. This technique is called time-domain thermoreflectance (TDTR), which is widely used in the fields of thermoelectrics and thermal engineering.[82-87] In TDTR, an ultrafast laser pulse heats the sample surface, and the time evolution of the temperature change is measured through the reflected light intensity using another laser pulse. Then, the TDTR signals are analyzed based on a heat conduction model to estimate thermal transport parameters. If the TDTR system is combined with an electromagnet to apply a magnetic field to a sample, it can be used to investigate MTR in magnetic materials and spintronic multilayers (Fig. 9).[56,57] Frequency-domain thermoreflectance is also commonly used in thermoelectrics and thermal engineering, where a sample is heated with an amplitude-modulated light and the thermal response is measured using another probe light.[88,89] The time-and frequency-domain thermoreflectance measurements are performed on a much smaller timescale than that for infrared thermography, enabling the determination of thermal conductivity and interfacial thermal conductance in thin film systems.

To quantify the temperature change accurately using the thermoreflectance methods, the thermoreflectance coefficient (∝ temperature derivative of reflectance) of the outermost surface of a sample must be large. Thus, a transducer layer with the given thermoreflectance coefficient, such as Al and Au, is usually formed on the sample surface to convert the reflectivity change into temperature information.

*3.4 Lock-in measurement with thermography and thermoreflectance*

The LIT and LITR methods are based on the combination of thermography and thermoreflectance thermometry with lock-in detection, respectively. These methods can be powerful tools for investigating thermal responses to excitations with high sensitivity, without attaching thermometers to a sample. In particular, lock-in detection is beneficial for thermoelectric studies in spin caloritronics, where a charge current is used as the input (Fig. 7). When a charge current is applied to a sample, not only thermoelectric cooling/heating but also Joule heating occurs. For the AC current, these contributions oscillate at different frequencies. If an AC current without a DC offset is applied to a sample, the thermoelectric contribution appears at the same frequency as the applied current, whereas the Joule heating contribution appears in DC or higher-harmonic signals [Fig. 10(b)].[36-38,69,70]

LIT and LITR measurements can be realized by synchronizing temperature measurement timing with an alternating input. For imaging measurements, thermal images are transferred to a computing unit, and the lock-in signal distribution, composed of the amplitude and phase delay, is extracted using time-domain Fourier analysis in each pixel of the images. For spot measurements, the Fourier analysis is performed using dedicated hardware, *i.e.*, a lock-in amplifier [Fig. 11(a)]. Table I summarizes the characteristics of the LIT and LITR methods.

To investigate the spin-caloritronic phenomena using the LIT and LITR methods, measurements of the magnetic field and lock-in frequency $f$ dependences are advantageous. As the spin-caloritronic phenomena depend on the **M** direction, their contributions to temperature modulation can be extracted by performing subtraction or addition of signals measured at positive, negative, and/or zero fields.[36-38,47] The $f$ dependence of the signals can be used to determine the thermoelectric or thermospin conversion efficiency and to reveal heat-source positions. By measuring the $f$ dependence of the temperature modulation signals and/or by comparing it with a heat diffusion model, the thermoelectric coefficients can be extracted.[48,49] The modeling of heat sources is crucial for this, which can be performed by measuring the $f$ dependence at high $f$ values for the following reason. When a single AC heat source oscillating at $f$ is generated, the temperature change is localized within a length scale determined by the thermal diffusion length $\sqrt{D/(\pi f)}$, where $D$ is the thermal diffusivity. Because the thermal diffusion length decreases with increasing $f$, the position and size of the heat source can be specified by the LIT and LITR measurements at high $f$ values.

Figure 10 schematically shows the setup and analysis procedures for the measurement of thermoelectric effects using the LIT method. A periodic charge current is applied to a sample using a current source, and thermal images of the sample surface are continuously acquired. The images are transferred to a computing unit and transformed into lock-in amplitude $A$ and phase-delay $\phi$ images via Fourier analysis. The $A$ ($\phi$) image shows the intensity (phase) distribution of the fundamental oscillation component of the temperature change synchronized with the charge current. The phase image contains information about the sign of the temperature change and time delay associated with thermal diffusion. If the effect of thermal diffusion is negligible, a phase delay of 0° (180°) indicates that heat is released (absorbed). The increased $\phi$ indicates that the temperature change is generated after a certain time; in the case of the time delay due to thermal diffusion, the $\phi$ values increase with increasing distance from heat sources. By combining the LIT system with an electromagnet, the magneto-thermoelectric and thermospin effects can be measured using the same procedures.

Figure 11(a) shows a schematic of the LITR system used to measure thermoelectric effects. A function generator is connected to the metal layer of the sample to apply an AC charge current. In this system, the sample surface is illuminated with visible light through a beam splitter, and the reflected light is detected using a photodetector. A laser or light-emitting diode can be used as a light source.[69,70] When the AC current causes the AC thermoelectric response, the reflected light intensity oscillates at the same frequency as the input current frequency. This optical signal is eventually converted into the amplitude and phase delay of the temperature modulation using the thermoreflectance coefficient. Figure 11(b) shows an example of the LITR measurement of SPE for a Pt/$Y_3Fe_5O_{12}$ junction system and AEE for a Ni film, where an Au transducer layer was formed by inserting an insulator layer to electrically separate the transducer from the Pt and Ni layers.[69] Temperature modulation signals induced by SPE and AEE with an amplitude of ~ 1 mK were clearly observed. The advantage of the LITR method is the measurement time scale; the maximum $f$ value in our setup exceeds 1 MHz, which is much larger than that for LIT (note that the lock-in frequency of LIT is limited by the frame rate of an infrared camera,[35] and is typically smaller than 100 Hz). This characteristic of LITR is useful for investigating the transient response of the magneto-thermoelectric and thermospin effects (Table I).

## 4. Heat Control Functionalities Enabled by Spin Caloritronics

Recently, we have clarified the detailed behaviors of various spin-caloritronic phenomena that output heat currents and demonstrated novel heat control functionalities mainly by using the LIT technique. Hereafter, by showing our representative experimental results, we introduce heat control functionalities unique to spin caloritronics.

*4.1 Local temperature modulation*

Figures 12(a) and 12(d) illustrate typical sample systems used for the LIT measurements of AEE and SPE, respectively.[37] The sample for the measurements of AEE consists of a ferromagnetic FePt film with a thickness of 10 nm epitaxially grown on a nonmagnetic $SrTiO_3$ (100) substrate. Here, the $L1_0$ ordering of the FePt film was controlled by the growth temperature, and the magnetization could be saturated with a relatively low magnetic field in both the perpendicular and in-plane directions. The sample for the measurements of SPE consists of a Pt film with a thickness of 10 nm grown on a ferrimagnetic $Y_3Fe_5O_{12}$ (111) substrate. In these samples, the FePt and Pt layers were processed into a U-shaped structure. To perform the LIT measurements, a square-wave-modulated AC current with zero DC offset was applied to the FePt and Pt films, where the current flows only in the metallic layers because $SrTiO_3$ and $Y_3Fe_5O_{12}$ are insulators. If AEE or SPE appears, the temperature change in linear response to the current is observed in the $A$ and $\phi$ images. By placing the U-shaped structure in the viewing area of thermal images, the dependence of the temperature change on the $J_c$ direction can be measured simultaneously, enabling a clear demonstration of the symmetries of AEE and SPE.

Figure 12(b) shows an example of the $A$ and $\phi$ images for the FePt film in the in-plane magnetized configuration, in which **M** of FePt is aligned in the $x$ direction by an external magnetic field.[37] No temperature change signal appears in the region where the **M** and $J_c$ directions are parallel to each other. On the other hand, a clear temperature change was observed in the region where the **M** and $J_c$ directions are perpendicular to each other, and $\phi$ of the signal changes by 180° with the $J_c$ reversal. This result indicates that the temperature changes with opposite signs occur in the right and left regions of the U-shaped structure. The sign of the temperature change in each region was found to be reversible by reversing the **M** direction. These behaviors are consistent with the symmetry of Eq. (4), and it can be concluded that the temperature change is due to the heat current in the $z$ direction driven by AEE.

Figure 12(e) shows the $A$ and $\phi$ images for the Pt/Y$_3$Fe$_5$O$_{12}$ system in the in-plane magnetized configuration in which **M** of Y$_3$Fe$_5$O$_{12}$ is aligned in the $x$ direction.[37] The Pt/Y$_3$Fe$_5$O$_{12}$ system exhibits temperature change signals with the same distribution as the AEE signals for the FePt film. Because Pt does not show AEE, the signal observed here is due to SPE driven by the spin current injection from Pt into Y$_3$Fe$_5$O$_{12}$. These experimental results are consistent with Eqs. (4)-(6); AEE and SPE exhibit the same symmetry in the in-plane magnetized configuration.

In Fig. 12(c) and 12(f), we show similar comparisons in the perpendicularly magnetized configuration, in which **M** of FePt and Y$_3$Fe$_5$O$_{12}$ is aligned in the direction perpendicular to the film surface.[37] The distribution of the current-induced temperature modulation for the FePt film in the perpendicularly magnetized configuration is completely different from that in the in-plane magnetized configuration; the temperature modulation appears at the edges of the entire FePt film, and the $\phi$ value changes by 180° between the inner and outer edges of the U-shaped structure, indicating that the heat currents flow in the in-plane directions satisfying the symmetry of Eq. (4). On the other hand, no temperature change was observed in the Pt/Y$_3$Fe$_5$O$_{12}$ system in the perpendicularly magnetized configuration, consistent with Eq. (6) showing that the SPE-induced heat current disappears when $\sigma \perp$ **M**. As shown here, the LIT method is a powerful tool for demonstrating the symmetries and functionalities of the thermoelectric and/or thermospin effects.

The LIT measurements of SPE revealed that a peculiar temperature distribution is generated by applying a spin current to the metal/magnetic material junction interface.[36] Figures 13(a) and 13(b) show the spatial distribution and profile across the charge current paths of the SPE-induced temperature change for the Pt/Y$_3$Fe$_5$O$_{12}$ system, respectively. Surprisingly, the temperature change was observed to be localized within the Pt/Y$_3$Fe$_5$O$_{12}$ junction where the spin current was injected. This temperature distribution is significantly different from the temperature change due to Joule heating, which is broadened from the heat-source positions through thermal diffusion [Figs. 13(d) and 13(e)]. The spatial distribution of the Joule-heating-induced temperature change depends strongly on $f$ because the temperature change takes a substantial amount of time to reach the steady state [Fig. 13(e)]. In contrast, the temperature change induced by SPE is independent of $f$ below 1 kHz for the Pt/Y$_3$Fe$_5$O$_{12}$ system [Figs. 11(b) and 13(b)], indicating that the SPE signals reach a steady state in a very short time.[36,69] These behaviors cannot be explained by a single heat source generation.

Daimon *et al.* showed that the unique temperature distribution accompanying SPE is reproduced by assuming that the spin current generates a dipolar heat source in the vicinity of the Pt/Y$_3$Fe$_5$O$_{12}$ interface.[36] The dipolar heat source consists of an adjacent pairing of a heat source ($+Q$) and a heat sink ($-Q$) of equal magnitude. The numerical simulations show that the dipolar heat sources placed near the surface of the system exhibit a temperature change localized within the scale of the heat source size, which is consistent with the experimental results for SPE [Fig. 13(c)]. To generate such a localized temperature change, the magnitude of the heat sources and sinks must be the same, *i.e.*, the net amount of heat must be macroscopically zero. If the net amount of heat is finite, a temperature change is generated even at positions far from the heat sources, as in the case of Joule heating [compare Fig. 13(c) and 13(f)]. Similar localized temperature distributions can also be obtained by AEE in thin films in the in-plane magnetized configuration[37,47] and by the Peltier effect in multilayer structures in the current-perpendicular-to-plane configuration.[63]

Dipolar heat sources are potentially applicable to pinpoint temperature controllers. Because such heat sources induce temperature changes localized on a nanometer to micrometer scale, target regions can be heated or cooled selectively without affecting the surrounding environment. Another merit of using dipolar heat sources is high-response heating/cooling because the temperature change induced by dipolar heat sources reaches a steady state within a very short time scale. To accomplish thermal management based on dipolar heat sources, not only materials showing large thermoelectric or thermospin conversion efficiency but also optimal thermal design is necessary. This is because the temperature change induced by dipolar heat sources strongly depends on the thermal resistance between the heat sources and sinks and on their sizes. Such dipolar heat source engineering is indispensable for establishing design guidelines for pinpoint temperature controllers (Fig. 1).

*4.2 Active control of thermoelectric conversion*

The magneto-thermoelectric and thermospin effects enable active thermal management. Because the direction and magnitude of heat currents generated by such effects are determined by magnetization, thermoelectric heating/cooling properties can be controlled in a reconfigurable manner through the interaction between magnetism and various degrees of freedom, such as mechanical strain, electric fields, and light. The reconfigurable control of thermoelectric conversion properties is impossible if only the conventional Peltier effect is used. Hereafter, we show proof-of-concept demonstrations of the active control of magneto-thermoelectric conversion.

*4.2.1 Strain-induced control of heat current*

It is known that strain application to magnetic materials can modulate magnetic anisotropy, and alter the **M** direction owing to the magnetoelastic or inverse magnetostrictive effect. Thus, strain-induced control of heat currents is possible by combining the magnetoelastic effect with the spin-caloritronic phenomena.

In 2021, Hirai *et al*. demonstrated uniaxial-strain-induced switching between cooling and heating generated by AMPE.[90)] As illustrated in Fig. 14(a), if a magnetic material has a negative magnetostrictive coefficient and a uniaxial tensile strain is applied along the initial **M** direction, the magnetic easy axis of the material can be rotated by 90°, resulting in the sign reversal of the AMPE-induced temperature modulation at the corners of the U-shaped structure [Eq. (1)]. The active control of AMPE by strain application was demonstrated using a U-shaped Ni film deposited on a flexible substrate and detected using the LIT technique. Figure 14(b) shows the $A_{\text{AMPE}}$ and $\phi_{\text{AMPE}}$ images for the Ni film, where $A_{\text{AMPE}}$ ($\phi_{\text{AMPE}}$) denotes the lock-in amplitude (phase) of the temperature modulation induced by AMPE. When **M** of the Ni film is along the transverse direction under a weak magnetic field, the heating (cooling) signal is generated at the left (right) corner [see the left images in Fig. 14(b)]. Applying the strain to the sample changed the $\phi_{\text{AMPE}}$ values on the corners by 180° without changing the magnetic field, indicating the strain-induced sign reversal of the AMPE signals [see the center images in Fig. 14(b)]. The strain-induced control of AMPE is reversible; $\phi_{\text{AMPE}}$ returns to the initial values after relaxing the strain [see the right images in Fig. 14(b)]. The observed behavior occurs in magnetic materials with negative magnetostrictive coefficients. The response of AMPE to strain application depends on the magnetoelastic properties of magnetic materials.

The combination of the magnetoelastic effect and AEE enables directional control of heat currents. In 2019, Ota *et al*. demonstrated the strain-induced 180° switching and 90° rotation of the heat currents due to AEE by using an in-plane magnetized Ni film and a perpendicularly magnetized TbFeCo film, respectively (Fig. 15).[91)] These functionalities are based on Eq. (4); when the magnetoelastic coupling changes the magnetic easy axis, the direction of the AEE-induced heat current is controlled in response to the change of the **M** direction.

*4.2.2 Electric-field-induced ON/OFF switching of heat current*

It is known that, when an electric field is applied to an ultrathin magnetic film, various magnetic parameters, such as the magnetic anisotropy, Curie temperature, and domain wall motion, are modulated. Thus, the electric field effect on magnetism has been investigated intensively to develop next-generation spintronic devices with low power consumption. By combining the electric field effect on magnetism with the spin-caloritronic phenomena, thermoelectric conversion and thermal transport properties can be controlled electrically without applying a magnetic field.

In 2019, Nakayama *et al*. demonstrated electric-field-induced on/off switching of heat currents due to AEE.[92)] If the ferromagnetic-to-paramagnetic phase transition is induced by an electric field effect, AEE-induced heat currents are eliminated under the electric field (Fig. 15). The AEE-induced heat currents can again be switched on by removing the electric field. To demonstrate this functionality, Nakayama *et al.* used an ultrathin Co film with a naturally oxidized surface and a solid-state capacitor structure, which exhibits a magnetic phase transition at room temperature due to electric-field-induced redox reactions at the interface between the Co and oxidized layers. Importantly, this electric-field-induced on/off switching of AEE appears only in the regions where the electric field is applied, enabling local control of thermoelectric cooling/heating. This is an advantage of the electric field effect that cannot be realized by the strain-induced effect discussed in the previous subsection. However, the electric field effect also has the disadvantage of being applicable only to ultrathin films because the electric field effect does not work for thick metals due to the screening effect.

*4.2.3 Magneto-optical design of heat current*

In 2020, Wang *et al.* proposed that the direction and distribution of heat currents generated by the spin-caloritronic phenomena can be changed simply by illuminating magnetic materials with visible light.[93] This functionality was demonstrated by combining AEE with all-optical helicity-dependent switching of magnetization. The all-optical helicity-dependent switching of magnetization can be used for magneto-optical recording because it enables magnetization reversal in magnetic materials as a result of circularly polarized light illumination without applying external magnetic fields, where the **M** direction is determined by the light helicity.[94-98] Importantly, the light-induced **M** reversal appears only in the illuminated area. Therefore, by patterning the illuminating light and determining its helicity, the distribution of the AEE-induced heat current and the resulting temperature change can be designed [Fig. 16(a)]. This approach enables not only pinpoint manipulation and flexible design of the heat current distribution but also on/off switching of the temperature change by illuminating the sample with linearly polarized light. Wang *et al.* demonstrated these functionalities by using perpendicularly magnetized Co/Pt multilayer films, which exhibit the highly efficient all-optical helicity-dependent switching of magnetization [Fig. 16(b)].[93] This versatile, reconfigurable, and reversible heat control paves the way for nanoscale thermal management. The concept of the magneto-optical design of heat currents can also be extended to various spin-caloritronic phenomena, such as SPE and AMPE, if in-plane magnetized materials are used.

*4.3 Spintronic thermal switching*

To make effective use of spintronic thermal management, the heat control functionalities must be integrated into the architecture of spintronic devices. Thus, MTR in magnetic multilayers is essential because it enables thermal switching in GMR and TMR devices. In this subsection, we summarize recently reported experiments on MTR.

In 2021, Nakayama *et al.* reported the observation of the giant MTR for the out-of-plane thermal conductivity in a fully epitaxial Cu/CoFe multilayer film.[57] To measure thermal conductivity using the TDTR method, the top surface of the multilayer film was covered by a thin Al transducer layer with a known thermoreflectance coefficient [Fig. 17(a)]. The TDTR measurements were performed in the front face heating/front face detection configuration with the application of an in-plane magnetic field $H$ (Fig. 9). As shown in the magnetization curve in Fig. 17(b), the Cu/CoFe multilayer film exhibits the parallel (antiparallel) magnetization configuration when $\mu_0 H > 80$ mT ($\mu_0 H \sim 0$ mT). Figure 17(b) also shows the $H$ dependence of the effective out-of-plane thermal conductivity of the Cu/CoFe multilayer film, estimated by analyzing TDTR signals using a heat conduction model.[57] The Cu/CoFe multilayer film was found to exhibit clear MTR; the thermal conductivity increases greatly with increasing $H$ and is saturated for $\mu_0 H > 80$ mT, *i.e.*, in the parallel magnetization configuration. The out-of-plane thermal conductivity change $\kappa_P - \kappa_{AP}$ and the MTR ratio ($\kappa_P - \kappa_{AP}$)/$\kappa_{AP}$ observed here reach 24.8 Wm$^{-1}$K$^{-1}$ and 150% at room temperature, respectively. The giant MTR for the Cu/CoFe multilayer film was found to appear even at temperatures of up to 400 K [Fig. 17(c)].

Figure 17(d) shows a comparison between MTR and GMR for the Cu/CoFe multilayer film.[57] In the temperature range from room temperature to 400 K, the MTR ratio was observed to be much larger than the magnetoresistance ratio in the current-perpendicular-to-plane configuration. This result cannot be explained by conventional spin-dependent electron transport, suggesting the contribution of additional heat carriers and mechanisms enhancing MTR. Magnons are a potential heat carrier that contributes to MTR, but their contribution is yet to be verified in magnetic multilayers. To confirm the magnon contribution in MTR, systematic measurements at low temperatures and high magnetic fields are necessary.

As reviewed here, giant MTR has recently been observed in magnetic multilayers. However, to achieve thermal management applications of MTR, the following must be addressed:

- Revealing microscopic mechanisms of giant MTR
- Finding materials and optimizing multilayer structures to further enhance the thermal conductivity change and MTR ratio
- Implementing nonvolatile and highly controllable spintronic thermal switching
- Observing MTR in TMR devices
- Demonstrating the operation of spintronic thermal switching in practical devices

We anticipate that rapid developments in heat detection techniques will further accelerate studies on MTR in magnetic and spintronic materials.

Notably, thermal conductivity of magnetic materials can also be modulated by an electric field. In 2020, Terakado *et al.* reported an example of such a functionality; they observed electric-field-induced modulation of thermal conductivity in spin-chain ladder cuprate films, where the magnon contribution to thermal conductivity plays an essential role.[99] The electric-field-induced control of thermal conductivity also has the potential to develop active thermal switching devices.

*4.4 Unidirectional remote heating*

In this subsection, we summarize the heat control functionalities based on magnetostatic spin waves, *i.e.*, the unidirectional spin-wave heat conveyer effect. Spin waves can be used to control remote heating by means of microwave excitation and the nonreciprocal nature of the magnetostatic surface spin waves. An *et al.* prepared a $Y_3Fe_5O_{12}$ slab mounted on a microwave antenna and measured the temperature distribution under microwave irradiation.[74] When the magnetostatic surface spin waves are not excited, microwave heating appears only near the antenna. In contrast, once the frequency of microwaves matches that of the magnetostatic surface spin waves, excited spin waves nonreciprocally transfer energy or heat to one edge of the slab, resulting in unidirectional heating of a distant area. Interestingly, the heating direction can be magnetically controlled because the propagation direction of nonreciprocal spin waves depends on the **M** direction.

While the basic behavior was revealed by the aforementioned work and subsequent studies,[74,75,100,101] the heat source distribution due to the unidirectional spin-wave heat conveyer effect remained to be clarified experimentally. To reveal the heat source distribution, Kainuma *et al.* performed LIT measurements for a rectangular $Y_3Fe_5O_{12}$ slab at high lock-in frequencies, where microwaves with chopped amplitude were used as the lock-in input [Fig. 18(a)].[102] As expected, by increasing *f*, the obtained LIT images clearly show sharp heating patterns owing to the suppression of temperature broadening caused by thermal diffusion [Fig. 18(b)]. By tuning the microwave frequency into the band of the magnetostatic surface spin waves, complicated heating patterns were observed on one side of the slab [Figs. 18(c) and 18(d)]. It was found that the heating pattern depends on the frequency of the excited spin waves, indicating that the heating is not due to simple damping but to the edge spin-wave dynamics.

The unique heat control functionalities of the unidirectional spin-wave heat conveyer effect come not only from the nonreciprocity but also from their long-range nature. The LIT measurements also confirmed that this effect can generate heat even through a macroscopic air gap. Because the dipole-dipole interaction, which governs the dynamics of the magnetostatic surface spin waves, is a long-range interaction, it can transfer energy over a gap. To demonstrate this functionality, Kainuma *et al.* used two $Y_3Fe_5O_{12}$ slabs separated by a distance ranging from 0 to 1.5 mm, with a microwave antenna attached near the end of one slab.[102] It was confirmed that heating signals appeared at the far end of the slab without the antenna. The observed heating pattern was found to be similar to that for a single $Y_3Fe_5O_{12}$ slab, indicating that the unidirectional spin-wave heat conveyer effect can be used even in the presence of mm-scale air gaps.

## 5. Material Screening for Spin Caloritronics

To apply the unique heat control functionalities of the spin-calorogenic phenomena practically, it is important to find materials with high spin-heat-charge current conversion performance. However, the spin-calorogenic phenomena that output heat currents have been investigated only in a limited number of materials. In this section, we show that the LIT technique is a powerful tool not only for elucidating the characteristics and mechanism of the spin calorogenics phenomena but also for searching materials.

The greatest merit of the LIT method is that the spatial distribution of the thermoelectric and thermospin conversion performance can be characterized within a viewing area of a thermal image. Therefore, by combining the LIT method with combinatorial materials science,[103,104] the composition dependence of the magneto-thermoelectric and thermospin effects can be obtained simultaneously. This is in sharp contrast to the conventional method, in which multiple samples with different compositions are prepared and characterized individually. The LIT method thus enables high-throughput investigations of the spin-calorogenic phenomena, contributing to finding materials with large spin-heat-charge current conversion coefficients. We demonstrated the validity of this high-throughput screening method using a composition-spread thin film formed on a single substrate by a combinatorial sputtering method.[105-107]

Figure 19 shows a successful demonstration of high-throughput screening for AEE in magnetic Heusler alloys $Co_2MnAl_{1-x}Si_x$.[107] Modak et al. fabricated a composition-spread $Co_2MnAl_{1-x}Si_x$ film on a single-crystalline MgO (001) substrate [Figs. 19(a) and 19(b)] and measured the spatial distribution of the temperature modulation signals induced by AEE and Joule heating using the LIT method. As shown in Figs. 19(c) and 19(e), the composition dependence of AEE can be obtained as continuous information from a single thermal image. In addition, the relative composition dependence of the electrical resistivity can be estimated from the spatial distribution of the temperature modulation induced by Joule heating [Figs. 19(d) and 19(f)].[107] The latter measurement is also useful, for example, to compare the thermoelectric power factors for different compositions.

Importantly, the high-throughput screening based on the LIT method and combinatorial thin film deposition is applicable to various spin-calorogenic phenomena. For instance, if a composition-spread metallic film is formed on $Y_3Fe_5O_{12}$ and SPE measurements are performed, the composition dependence of the spin Hall effect can be investigated in a single sample.[105,106] By comprehensively screening the thermoelectric and thermospin conversion properties by this method and performing detailed characterization only for promising compositions, it is possible to develop good spin-calorogenic materials efficiently. It is also important to incorporate materials informatics methods into spin calorogenics based on accumulated data. We anticipate that materials science for spin calorogenics will be invigorated and that spin-heat-charge current conversion performance will be further improved in the near future.

## 6. Thermopile Structures for Magneto-Thermoelectric Effects

To enhance the heating/cooling power of the magneto-thermoelectric effects, it is necessary both to improve the thermoelectric conversion coefficients and to design device structures. In conventional Peltier modules, many thermocouples comprising p-type and n-type semiconductors are connected electrically in series and thermally in parallel, where the heating/cooling power is proportional to the number of thermocouples. Such a device structure is called a thermopile. A thermopile is also effective for increasing the heating/cooling power of AMPE and AEE. However, the thermopile structures for the magneto-thermoelectric effects are conceptually different from those of conventional Peltier modules.

*6.1 Thermopile structure for anisotropic magneto-Peltier effect*

The design guideline for an AMPE-based thermopile is simple. The AMPE-induced heating/cooling power can be enhanced by closely arranging the AMPE elements, *i.e.*, ferromagnetic wires with curved/bent structures and/or non-uniform magnetization configurations. If the AMPE elements exhibit a temperature change with the same sign, the total heating/cooling power is proportional to the number of integrated elements. The simplest AMPE element is an L-shaped ferromagnetic wire with an inner angle of 90°; when the element is uniformly magnetized along one of the legs of the L shape, the AMPE-induced temperature modulation is maximized at the corner due to the symmetry of AMPE [Eq. (1)]. Thus, one of the basic thermopile structures for AMPE is a cross-shaped thermopile comprising four L-shaped elements, where the total heating (cooling) power is +4$Q$ (−4$Q$) with $Q$ being the heat generated from a single wire [Fig. 20(a)]. To generate a large temperature change, it is necessary to connect the corners of the L-shaped elements thermally. The total heating/cooling power can be further increased by stacking the cross-shaped thermopiles three-dimensionally. The same argument can be applied to a zigzag-shaped thermopile [Fig. 20(b)]. Although the cross-shaped thermopile is suitable for local temperature control, the zigzag-shaped thermopile is applicable to temperature control over a wide area despite its low integration density. The AMPE-based thermopiles are workable even if the angle of the elements is not 90°. However, in this case, the magnetization distribution should be appropriately designed with the aid of shape and/or crystal magnetic anisotropies and exchange bias.

To demonstrate the effectiveness of the above concept, we constructed a prototypical cross-shaped thermopile comprising four L-shaped polycrystalline Ni wires and performed LIT measurements.[108] In this thermopile, the corners of the Ni wires are thermally connected through a thermal adhesive attached around the center. When the thermopile was uniformly magnetized and the charge current direction in each wire was appropriately set, AMPE-induced temperature modulation was observed around the center of the thermopile with a magnitude much larger than that for a single wire (Fig. 21).[108] We note that the anisotropy of the Peltier coefficient of Ni is only ~ 2%. By constructing the AMPE-based thermopile using magnetic materials with larger $\Delta\Pi$, the temperature change can be further enhanced.

*6.2 Thermopile structure for anomalous Ettingshausen effect*

The orthogonal relation between the input charge current and output heat current in AEE allows the construction of thermopiles with much simpler structures than conventional Peltier modules [Eq. (4)]. The basic structures of AEE-based thermopiles are shown in Fig. 22. When magnetic materials exhibiting AEE are stacked to generate heat currents in the same direction in response to an applied charge current and the layers are thermally connected, the temperature change output increases in proportion to the number of layers, *i.e.*, the total length of the thermopile along the heat current. If the layers are uniformly magnetized, the charge current must be applied in the same direction to align the heat current direction, which complicates the wiring for a serial connection [Fig. 22(a)]. This problem can be overcome by alternately stacking two different magnetic materials, A and B, with different $\Pi_{AEE}$. In this structure, even if the wiring is simplified and the charge current direction in the material B is opposite to that in the material A, the total temperature change can be enhanced, where the enhancement ratio depends on the difference in $\Pi_{AEE}$ between the magnetic materials A and B [Fig. 22(b)]. However, the advantage of the transverse thermoelectric effect partially disappears when the junction structures comprising different materials are used. A good enhancement ratio is obtained when $\Pi_{AEE} > 0$ ($\Pi_{AEE} < 0$) for the material A (B), while magnetic materials with negative anomalous Ettingshausen coefficients are limited (note that Fe and a $Nd_2Fe_{14}B$-type magnet exhibit small negative $\Pi_{AEE}$ values[40,48]). Therefore, the thermopile structure illustrated in Fig. 22(b) is not useful. To construct the AEE-based thermopile with a single material and simplified wiring, the **M** direction of each layer must be alternately reversed [Fig. 22(c)]. Such a magnetization configuration can be achieved by designing the magnetic anisotropy and coercivity and/or by using permanent magnets.

## 7. Conclusion

In this review article, we introduced the concept of spintronic thermal management as a new direction in spin caloritronics. To demonstrate the essence of this concept, we summarized the basic principles, behaviors, and unique heat control functionalities of the spin-caloritronic phenomena that output heat currents. Most of the phenomena reviewed here are measurable owing to recent developments in heat detection techniques. Although spintronic thermal management is still in its infancy in terms of both fundamental physics and applications, the final goal is to improve the efficiencies of electronics and spintronics devices through thermal management. The heat control functionalities introduced in this article are based on the intrinsic physical properties of magnetic and spintronic materials and, thus, are highly compatible with existing electronic and spintronic devices. Importantly, the spin-caloritronic phenomena appear over a wide temperature range, including room temperature.

An important task in the study of spintronic thermal management is to discover new principles and materials that enable highly efficient spin-heat-charge current interconversions. For example, in the case of AEE, the dimensionless figure of merit is still in the order of $10^{-3}$ to $10^{-4}$,[25,40,48,49] and it is necessary to improve the thermoelectric conversion performance significantly. If $\Pi_{AEE}$ is improved by an order of magnitude from the current record-high value and the thermal conductivity is reduced by half by nanostructuring,[5,6] the figure of merit for AEE would exceed 0.1. With this improvement, new thermoelectric applications may emerge owing to the versatility and functionality of AEE, even if the figure of merit is lower than that for the Peltier effect. As discussed in Sect. 2.1.3, the use of thermoelectric/magnetic hybrid materials might be a breakthrough for improving the performance of magneto-thermoelectric conversion. To realize applications of spintronic thermal management, more systematic and comprehensive investigations of the spin-caloritronic phenomena using various classes of materials are indispensable, and the high-throughput measurement techniques discussed in Sect. 5 will become increasingly important, as will machine learning analyses.[109,110]

This review article includes only a specific aspect of spintronic thermal management but many new research trends are emerging. In the field of thermoelectrics, the effect of magnons or spin fluctuations on the Seebeck effect, *i.e.*, the magnon drag effect, has attracted increasing attention as a principle to improve the Seebeck coefficient.[111-113] Although this article focuses only on linear-response transport phenomena, investigations on nonlinear spin caloritronics have recently been reported.[114] By expanding the physics of spin caloritronics, unconventional thermoelectric effects based on dielectric polarization transport have been predicted theoretically.[115-117] As exemplified by these research trends, spintronic thermal management is essentially interdisciplinary. We anticipate that basic and applied studies on spintronic thermal management will be further developed by introducing various principles of condensed matter physics and technologies of thermal energy engineering into spin caloritronics.

**Acknowledgment**

The experimental results shown in this review article were obtained from the collaboration with T. An, H. Awano, D. Chiba, S. Daimon, R. Das, K. Hasegawa, B. Hillebrands, T. Hioki, T. Hirai, R. Itoh, S. Iwamoto, Y. Kainuma, T. Koyama, A. Miura, R. Modak, H. Nagano, H. Nakayama, T. Ohkubo, S. Ota, K. Oyanagi, D. Prananto, E. Saitoh, Y. Sakuraba, T. Seki, H. Sepehri-Amin, J. Shiomi, Y. K. Takahashi, K. Takanashi, P. V. Thach, J. Uzuhashi, V. I. Vasyuchka, J. Wang, B. Xu, and T. Yamazaki. The authors thank many collaborators for valuable discussions and group members for preliminary reviews of the manuscript. This work was mainly supported by CREST "Creation of Innovative Core Technologies for Nano-enabled Thermal Management" (JPMJCR17I1) from JST, Japan. The authors also appreciate the contributions of T. An, Y. Miura, and H. Nagano to this project.

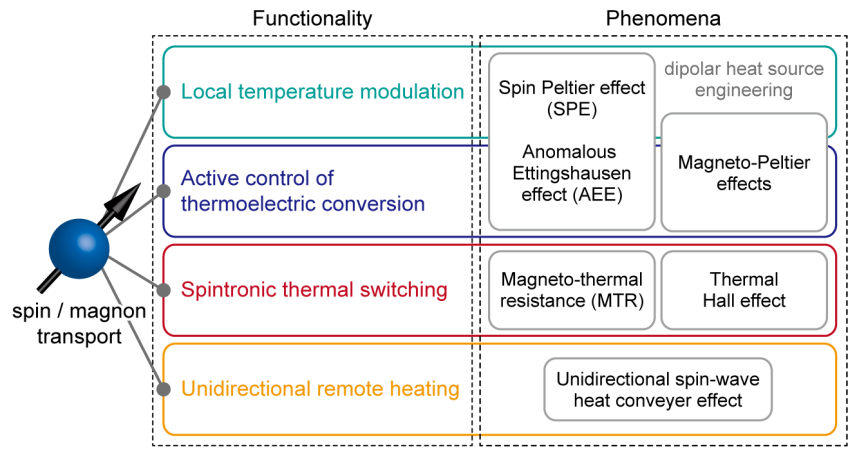

**Fig. 1.** Concept of spintronic thermal management.

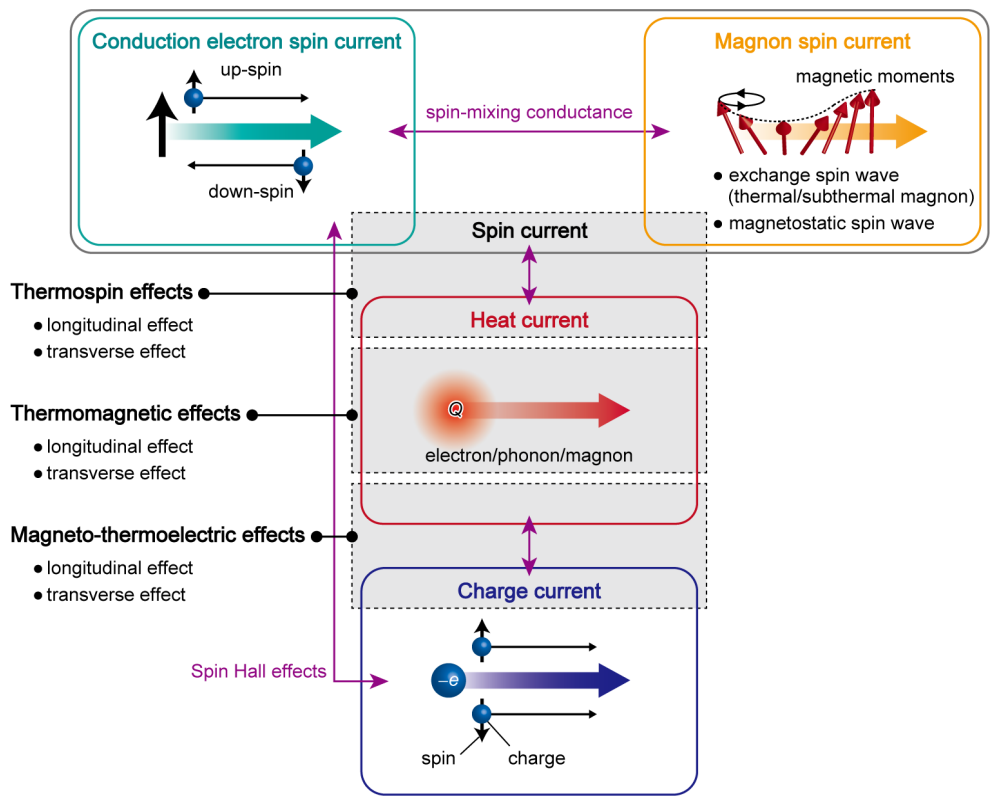

**Fig. 2.** Conversion phenomena between the charge current, heat current, conduction electron spin current, and magnon spin current. The categorization of the magneto-thermoelectric effects, thermomagnetic effects, and thermospin effects is marked with gray areas. $-e$ and $Q$ denote the electron charge and heat, respectively.

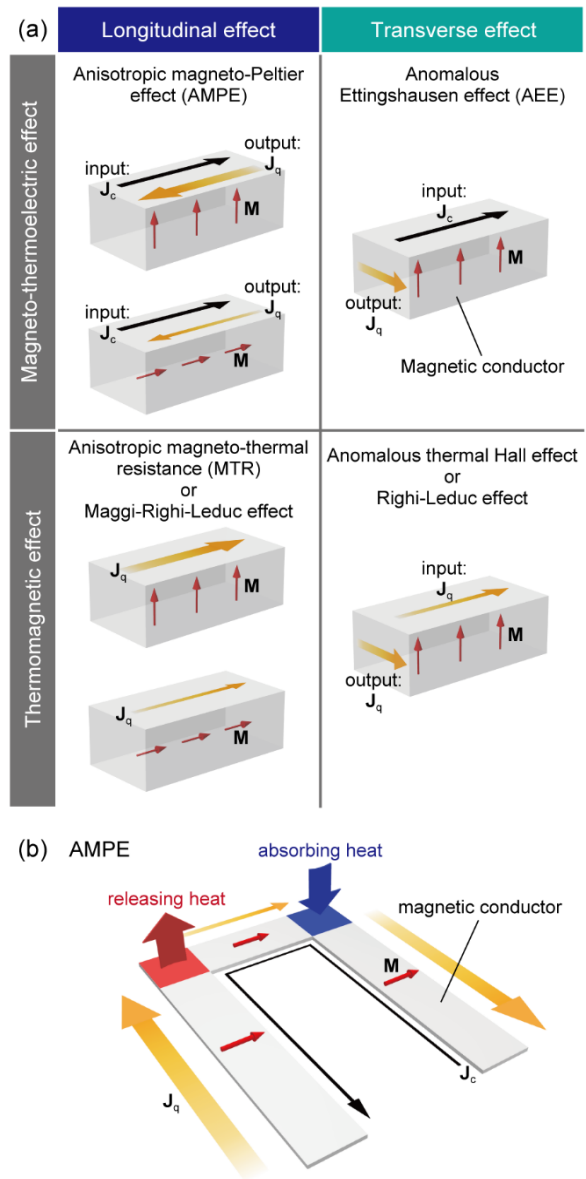

**Fig. 3.** (a) Schematics of the longitudinal/transverse magneto-thermoelectric and thermomagnetic effects, *i.e.*, the anisotropic magneto-Peltier effect (AMPE), anomalous Ettingshausen effect (AEE), anisotropic magneto-thermal resistance (MTR) or Maggi-Righi-Leduc effect, and anomalous thermal Hall effect or Righi-Leduc effect. (b) Schematic of the sample structure typically used for measuring AMPE. **J**$_c$, **J**$_q$, and **M** denote the charge current, heat current, and magnetization vector, respectively.

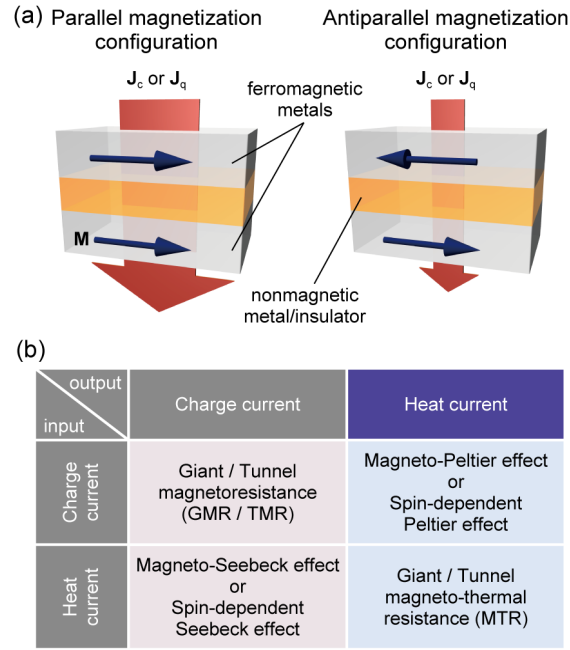

**Fig. 4.** (a) Schematics of charge and heat current transports in a multilayer comprising two ferromagnetic metals separated by a nonmagnetic metal (insulator), *i.e.*, a giant magnetoresistance (GMR) [tunnel magnetoresistance (TMR)] device, in the parallel and antiparallel magnetization configurations. (b) Charge, heat, and thermoelectric transport phenomena in GMR and TMR devices.

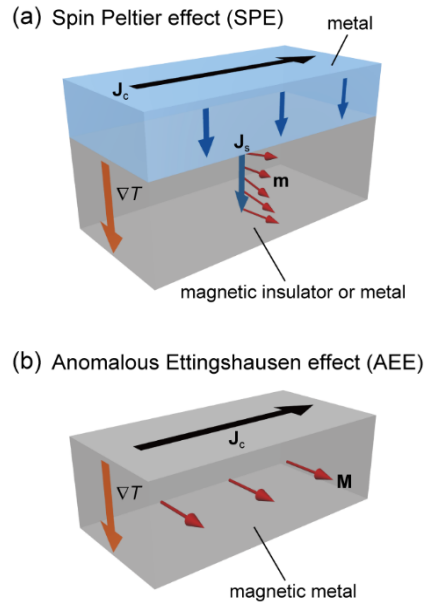

**Fig. 5.** (a) Schematic of the spin Peltier effect (SPE) induced by the spin Hall effect in a junction structure comprising a metal exhibiting spin-orbit interaction and a magnetic insulator or metal exhibiting magnon transport. (b) Schematic of AEE in a magnetic metal in the in-plane magnetized configuration. $\mathbf{J}_s$, $\mathbf{m}$, and $\nabla T$ denote the spatial direction of the spin current, magnetic moment, and temperature $T$ gradient, respectively.

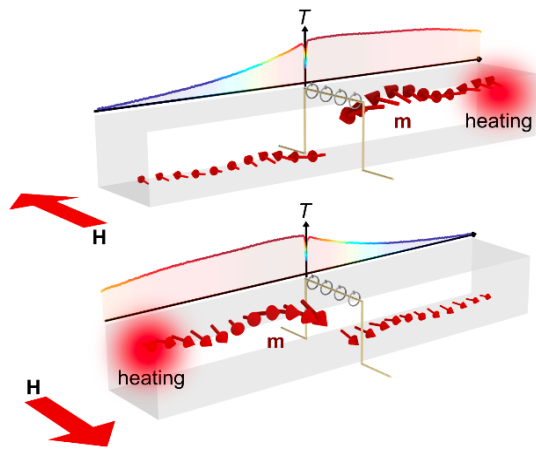

**Fig. 6.** Schematic of the unidirectional spin-wave heat conveyer effect excited by a microwave antenna located at the center of a magnetic material. Magnetostatic surface spin waves, collective dynamics of **m**, are excited near the antenna and propagate only in one direction specified by the nonreciprocal nature. This induces $\nabla T$ as a result of the remote heating. The direction of $\nabla T$ is reversed when **M** is reversed by the magnetic field **H**.

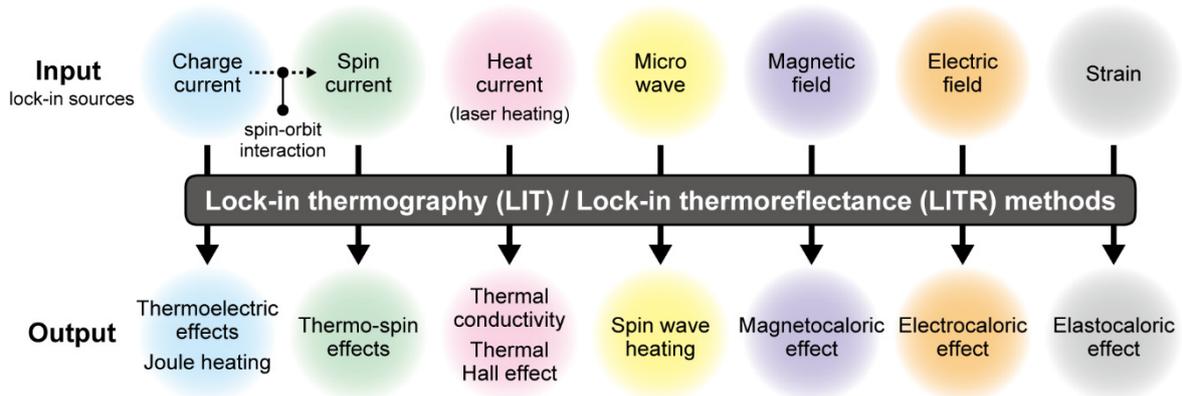

**Fig. 7.** Applications of the lock-in thermography (LIT) and lock-in thermoreflectance (LITR) methods.

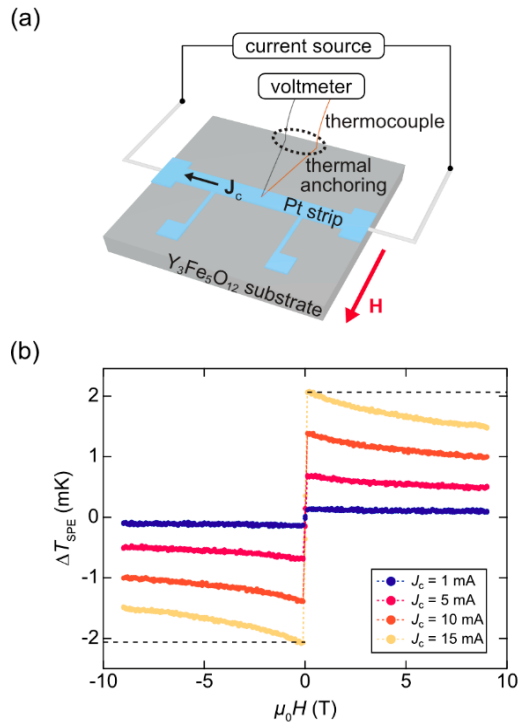

**Fig. 8.** (a) Schematic of the experimental setup for measuring SPE in a Pt/Y$_3$Fe$_5$O$_{12}$ system by a thermocouple. The junction point of the thermocouple was placed on the Pt strip covered by a thin insulation layer and the thermocouple was tightly connected using varnish. The output voltage of the thermocouple was monitored with a nanovoltmeter synchronized with a precision current source connected to the Pt strip, where the difference in the voltage $\Delta V_{TC}$ in response to the sign reversal of the charge current was measured. To control the **M** direction of Y$_3$Fe$_5$O$_{12}$, **H** with magnitude $H$ was applied in the direction perpendicular to the Pt strip. (b) $H$ dependence of the SPE-induced temperature change $\Delta T_{SPE}$ on the surface of the Pt/Y$_3$Fe$_5$O$_{12}$ system, estimated from the $\Delta V_{TC}$ values, for various values of the charge current applied to the Pt strip $J_c$. To obtain $\Delta T_{SPE}$, the $H$-odd component of the temperature change was extracted. The details of the experiments are shown in Ref. 68.

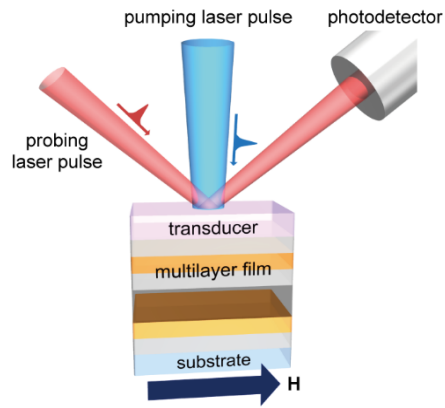

**Fig. 9.** Schematic of the time-domain thermoreflectance (TDTR) measurement for MTR in a magnetic multilayer film.

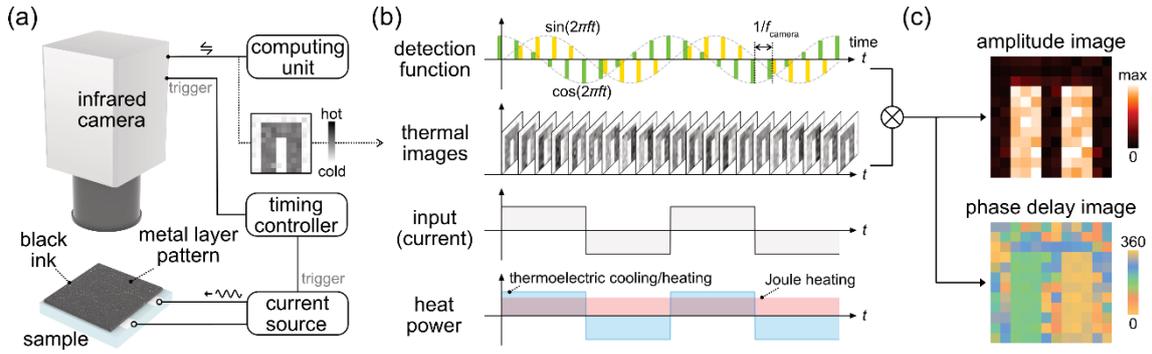

**Fig. 10.** (a) Schematic of an experimental setup for the LIT measurements of the thermoelectric and thermospin effects. A current source is connected to the metal layer of a sample. The infrared camera captures the surface temperature of the sample, coated with black ink, at the timing synchronized with an AC charge current applied to the metal layer. The thermal images are continuously transferred to a computing unit and Fourier analysis is performed using detection functions. (b) Time $t$ charts for the detection functions, thermal images, input current, and output heat power in the LIT measurements. When a square-wave-modulated AC charge current with zero DC offset is applied to the sample, the heat induced by the thermoelectric and thermospin effects oscillates with the same frequency as the current, whereas the Joule heating contribution is constant in time. $f$ and $f_{camera}$ denote the lock-in frequency and the frame rate of an infrared camera, respectively. (c) Lock-in amplitude and phase delay images obtained from the Fourier analysis.

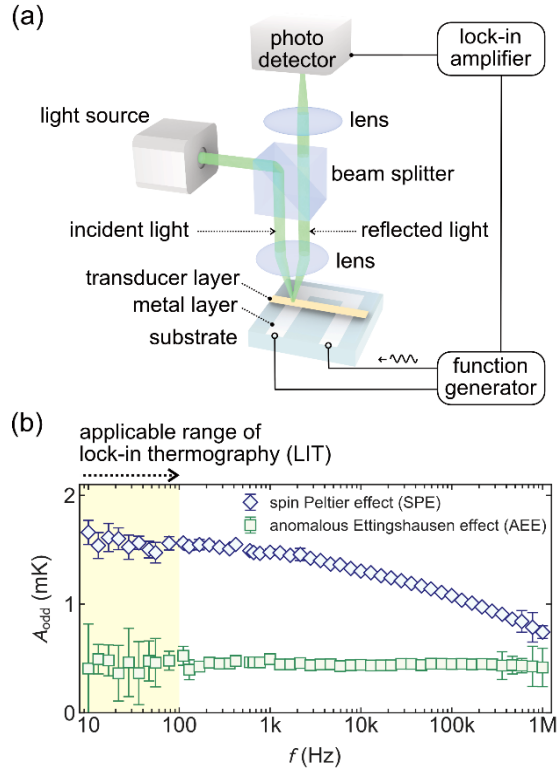

**Fig. 11.** (a) Schematic of an experimental setup for the LITR measurements of the thermoelectric and thermospin effects. The metal layer of a sample is connected to a function generator and covered by a transducer layer. Visible light is introduced to the surface of the transducer layer through a beam splitter and then the reflected light intensity is focused to the photodetector. The AC photodetector signals synchronized with the AC current applied to the metal layer are measured with a lock-in amplifier, which provides temperature information in the same manner as LIT. (b) $f$ dependence of the lock-in amplitude showing the $H$-odd dependence $A_{odd}$ for the Pt/$Y_3Fe_5O_{12}$ system and Ni film, measured by the LITR method. The samples were covered by Au transducer layers through insulating barriers. The $A_{odd}$ signals for the Pt/$Y_3Fe_5O_{12}$ system (Ni film) are attributed to SPE (AEE). The details of the experiments are shown in Ref. 69.

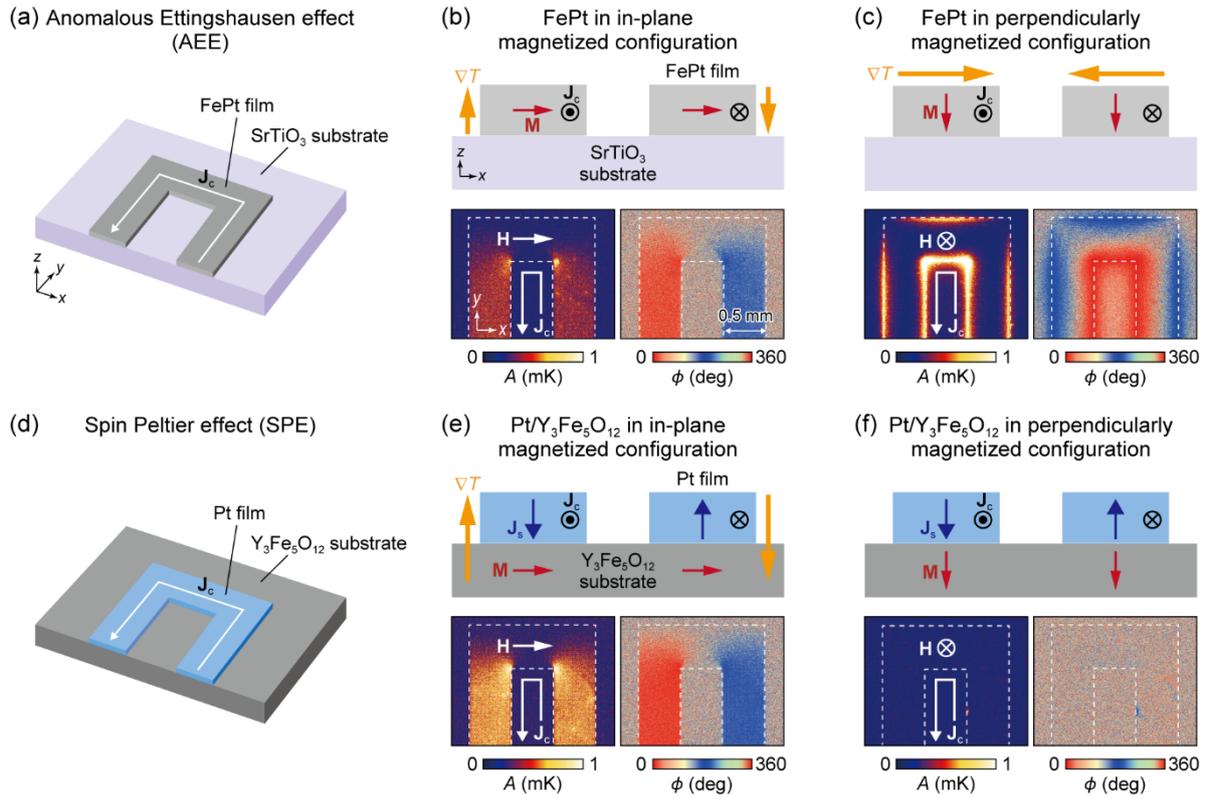

**Fig. 12.** (a) Schematic of the FePt/SrTiO$_3$ sample used for the LIT measurements of AEE. (b),(c) Schematics of the symmetry of AEE and the lock-in amplitude $A$ and phase $\phi$ images for the FePt/SrTiO$_3$ sample in the in-plane magnetized and perpendicularly magnetized configurations. (d) Schematic of the Pt/Y$_3$Fe$_5$O$_{12}$ system used for the LIT measurements of SPE. (e),(f) Schematics of the symmetry of SPE and the $A$ and $\phi$ images for the Pt/Y$_3$Fe$_5$O$_{12}$ system in the in-plane magnetized and perpendicularly magnetized configurations. During the LIT measurements, a square-wave-modulated AC charge current with $f = 25$ Hz, $J_c = 10$ mA, and zero DC offset was applied to the FePt and Pt layers. The details of the experiments are shown in Ref. 37.

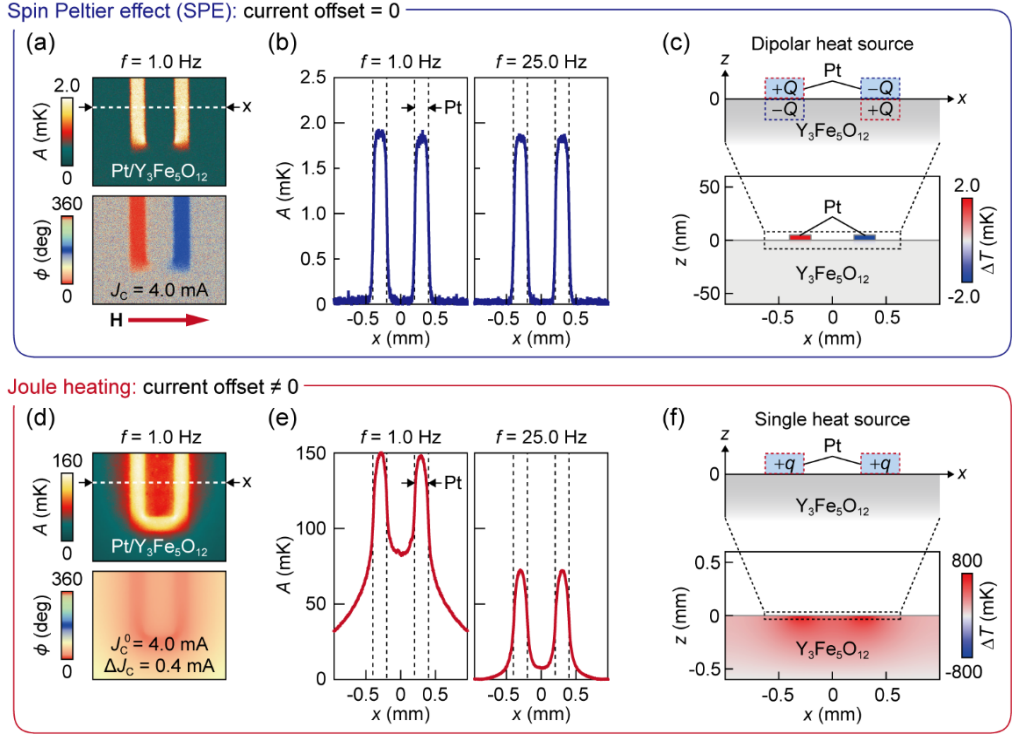

**Fig. 13.** (a) $A$ and $\phi$ images for the Pt/$Y_3Fe_5O_{12}$ system at $\mu_0 H = 20$ mT and $f = 1.0$ Hz, measured when the square-wave-modulated AC charge current with the amplitude $J_c = 4.0$ mA and zero offset was applied to the U-shaped Pt film. In this condition, the SPE signals free from Joule heating can be observed. (b) SPE-induced $A$ profiles along the $x$ direction across the Pt films at $f = 1.0$ Hz (left) and 25.0 Hz (right). (c) Calculated temperature difference $\Delta T$ distributions induced by the dipolar heat sources on the Pt/$Y_3Fe_5O_{12}$ interfaces. (d) $A$ and $\phi$ images for the Pt/$Y_3Fe_5O_{12}$ system at $\mu_0 H = 0$ mT and $f = 1.0$ Hz, measured when the square-wave-modulated AC charge current with the amplitude $\Delta J_c = 0.4$ mA and DC offset $J_c^0 = 4.0$ mA was applied to the Pt film. In this condition, the temperature modulation signals are dominated by the Joule heating contribution. (e) Joule-heating-induced $A$ profiles along the $x$ direction across the Pt films at $f = 1.0$ Hz (left) and 25.0 Hz (right). (f) Calculated $\Delta T$ distributions induced by the single heat sources on the Pt film of the Pt/$Y_3Fe_5O_{12}$ system. The details of the experiments and numerical calculations are shown in Ref. 36.

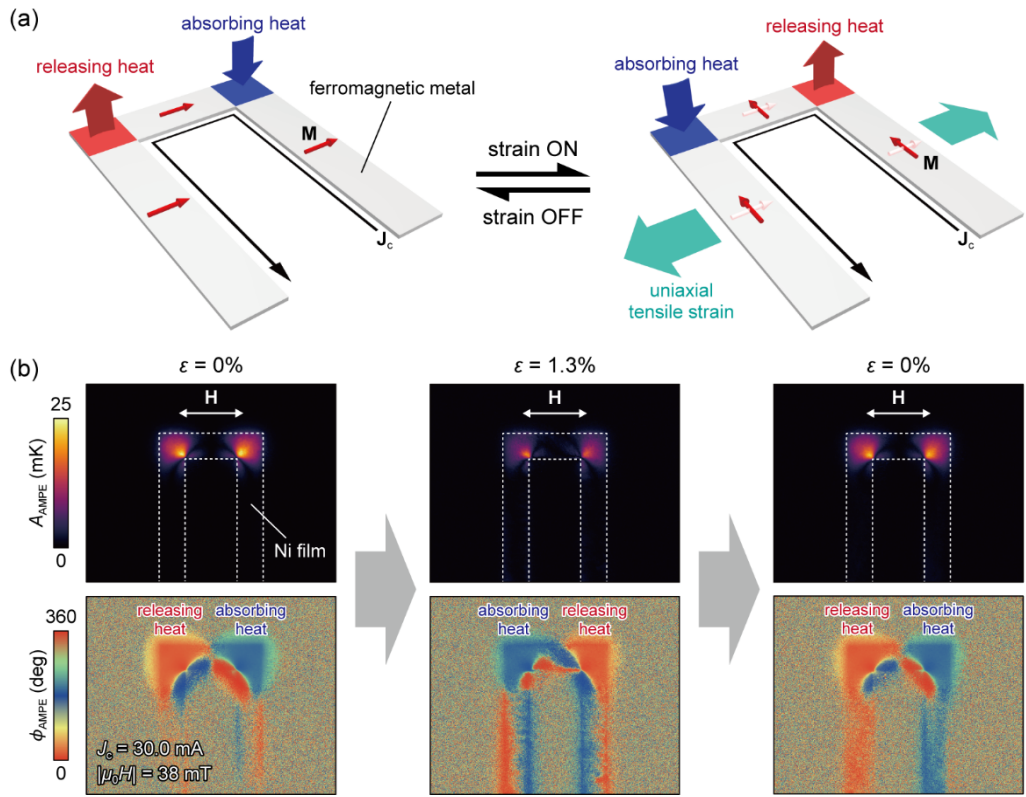

**Fig. 14.** (a) Schematics of strain-induced heating/cooling switching induced by AMPE. (b) Lock-in amplitude $A_{\text{AMPE}}$ and phase $\phi_{\text{AMPE}}$ images of the AMPE-induced temperature modulation for the U-shaped Ni film on the flexible substrate at $|\mu_0 H| = 38$ mT and $J_c = 30.0$ mA, measured when the magnetic field was applied along the transverse direction. When the in-plane uniaxial tensile strain $\varepsilon$ was applied to the sample, the sign of the AMPE-induced temperature modulation was reversed (center images). By releasing the strain, the sign of the temperature modulation goes back to the initial state (right images). The details of the experiments are shown in Ref. 90.

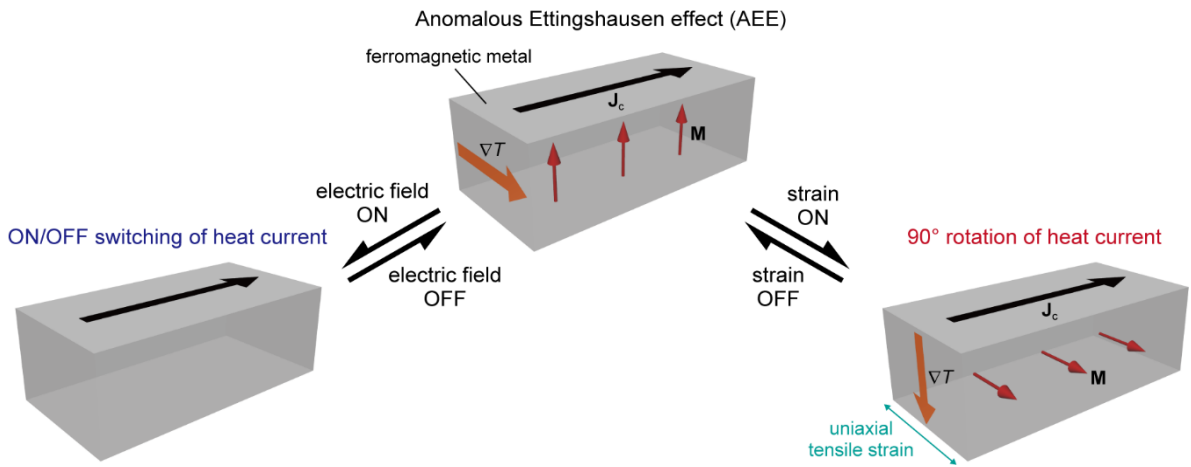

**Fig. 15.** Schematic of active control of AEE by an electric field or strain application.

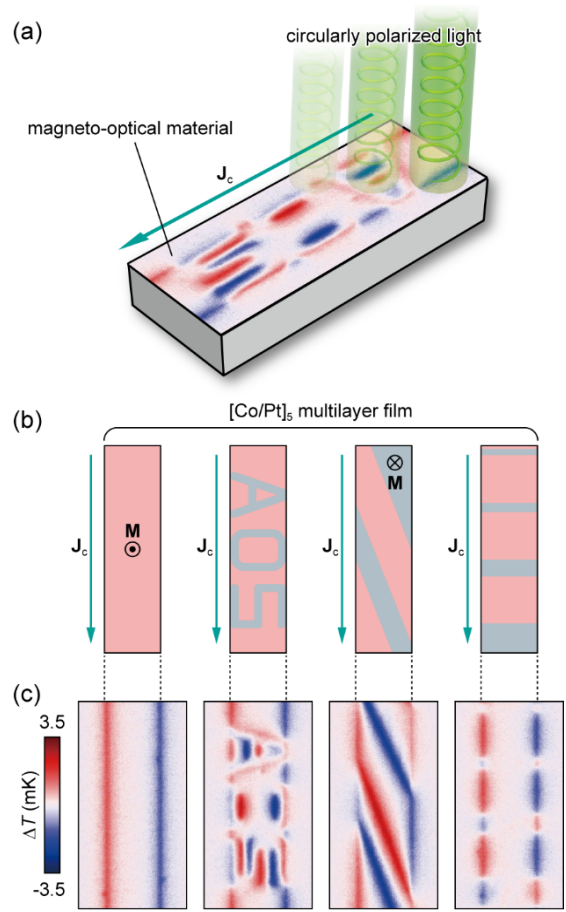

**Fig. 16.** (a) Magneto-optical design of the AEE-induced heat current and resultant temperature distribution. (b) Schematics of the spatial distribution of magnetic domains predesigned via the all-optical helicity-dependent switching of magnetization. (c) AEE-induced temperature modulation $\Delta T$ for the [Co/Pt]$_5$ multilayer films with the magnetic domain patterns shown in (b), measured without applying a magnetic field. The subscript 5 denotes the number of the Co/Pt bilayers. $\Delta T$ was measured by the LIT method through the relation $A\cos\phi$, which shows the current-induced temperature modulation with the sign information. The details of the experiments are shown in Ref. 93.

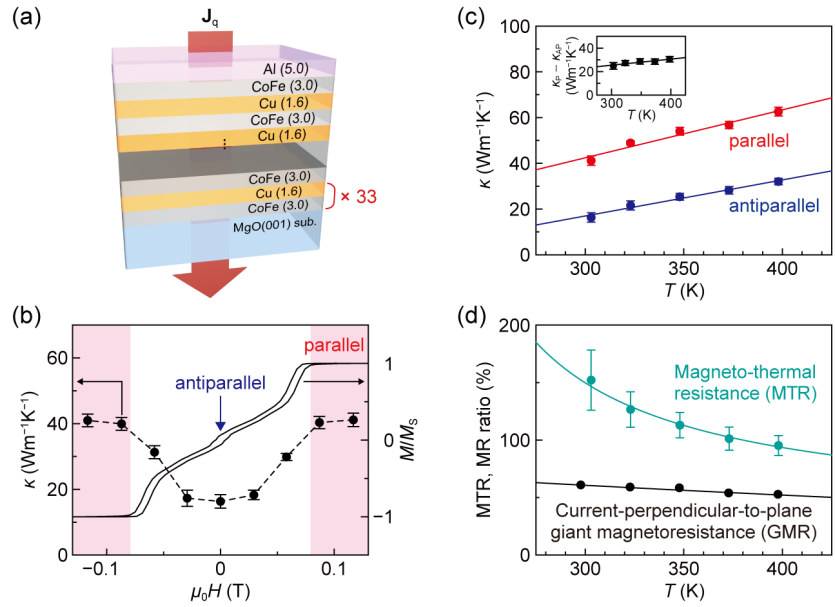

**Fig. 17.** (a) Schematic of the Cu/CoFe multilayer film used for the TDTR measurements. The Cu/CoFe multilayer film was covered by an Al transducer layer. (b) $H$ dependence of the effective out-of-plane thermal conductivity $\kappa$ and the normalized magnetization $M/M_s$ of the Cu/CoFe multilayer film at room temperature. $M_s$ denotes the saturation magnetization. (c) $T$ dependence of $\kappa$ of the Cu/CoFe multilayer film in the parallel and antiparallel magnetization configurations. The inset shows the $T$ dependence of $\kappa_P - \kappa_{AP}$, where $\kappa_P$ ($\kappa_{AP}$) is the thermal conductivity in the parallel (antiparallel) magnetization configuration. (d) $T$ dependence of the MTR and magnetoresistance (MR) ratios for the Cu/CoFe multilayer films in the current-perpendicular-to-plane configuration. The details of the experiments and analyses are shown in Ref. 57.

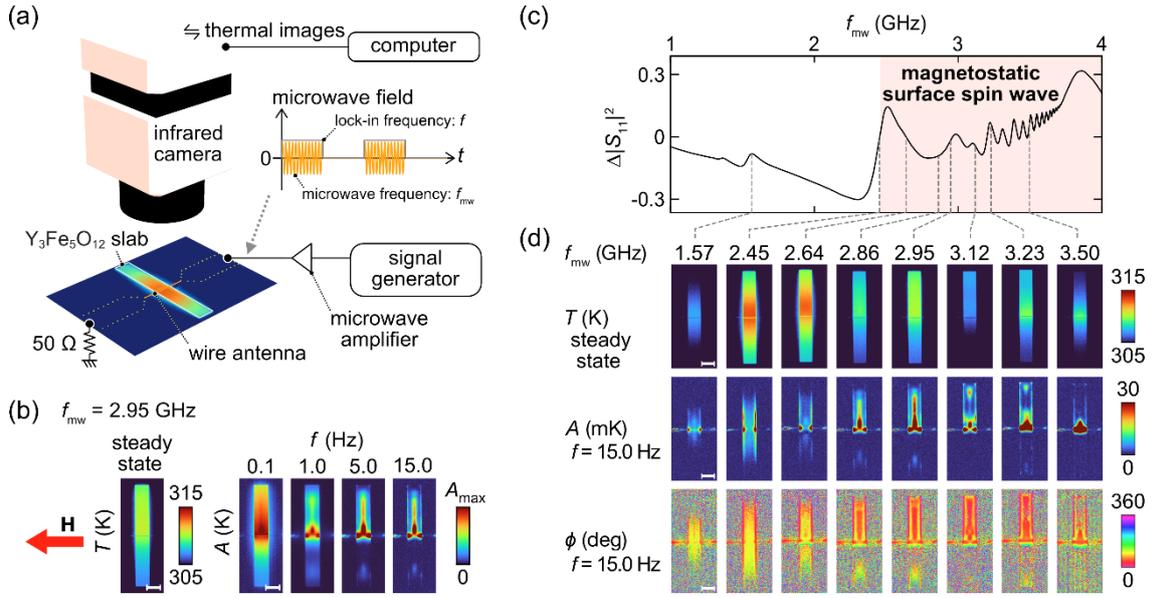

**Fig. 18.** (a) Schematic of the experimental setup for the LIT measurements of the unidirectional spin-wave heat conveyer effect. The infrared camera is synchronized to a signal generator, which applies microwaves with the field frequency $f_{mw}$ to the antenna on the $Y_3Fe_5O_{12}$ slab. The power of the microwaves is chopped at the lock-in frequency $f$. The antenna is also connected to a 50 Ω terminator. (b) Comparison of the steady-state $T$ image (left) and $A$ images (right) for various values of $f$ for the $Y_3Fe_5O_{12}$ slab under the excitation of the magnetostatic surface spin waves. The temperature modulation distribution in the $A$ images becomes sharper with increasing $f$, clarifying the heat source positions induced by the unidirectional spin-wave heat conveyer effect. $A_{max}$ on the color bar represents the maximum values in the $A$ images, where $A_{max}$ = 1.00, 0.30, 0.10, and 0.04 K for $f$ = 0.1, 1.0, 5.0, and 15.0 Hz, respectively. The length of the white scale bar is 2.0 mm. (c) Microwave reflection spectrum change $\Delta|S_{11}|^2$, measured with a vector network analyzer without a microwave amplifier. The magnetostatic surface spin waves are excited in the $Y_3Fe_5O_{12}$ slab when $f_{mw}$ > 2.5 GHz. (d) $T$, $A$, and $\phi$ images for various values of $f_{mw}$. The LIT images were obtained at $f$ = 15.0 Hz. The details of the experiments are shown in Ref. 102.

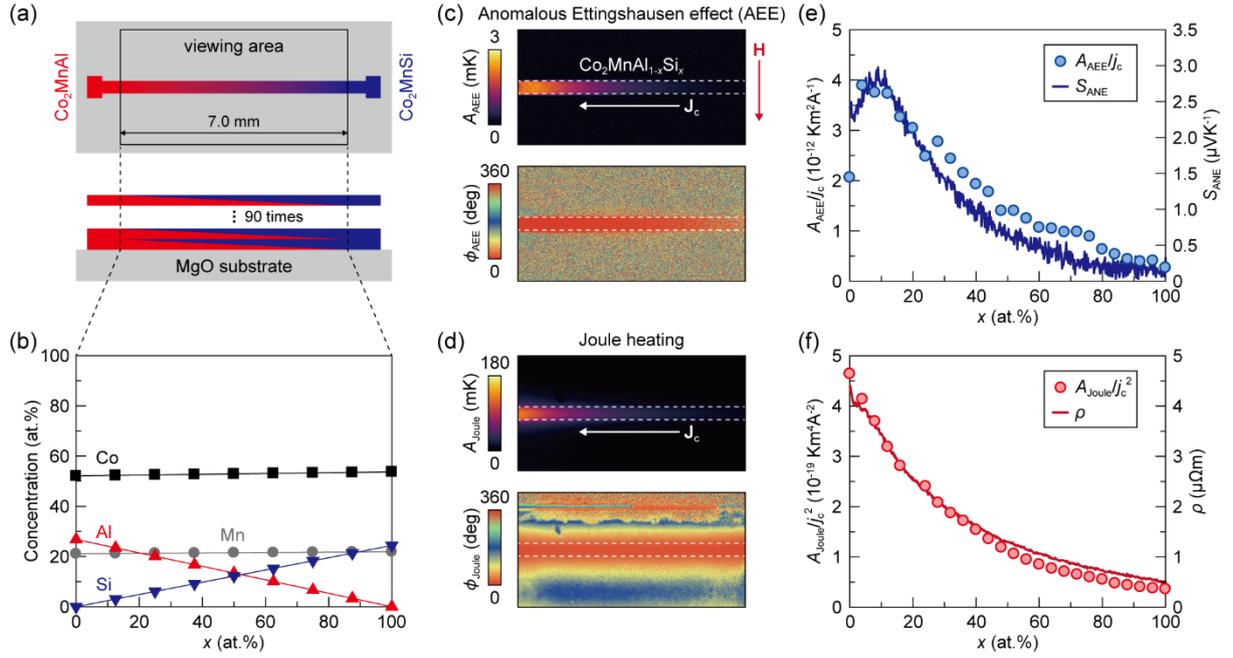

**Fig. 19.** (a) Schematic of the composition-spread $Co_2MnAl_{1-x}Si_x$ film. The thickness of the composition-spread film was fixed at ~ 50 nm at every position. (b) $x$ dependence of the Co, Mn, Al, and Si concentrations in the $Co_2MnAl_{1-x}Si_x$ film. (c),(d) Lock-in amplitude $A_{AEE}$ ($A_{Joule}$) and phase $\phi_{AEE}$ ($\phi_{Joule}$) images of the temperature modulation induced by AEE (Joule heating) for the $Co_2MnAl_{1-x}Si_x$ film at $f$ = 25.0 Hz. (e) $x$ dependence of $A_{AEE}/j_c$ and the anomalous Nernst coefficient $S_{ANE}$. $j_c$ denotes the square-wave amplitude of the charge current density applied to the $Co_2MnAl_{1-x}Si_x$ film. (f) $x$ dependence of $A_{Joule}/j_c^2$ and the electrical resistivity $\rho$. $S_{ANE}$ and $\rho$ were measured directly at several positions on the $Co_2MnAl_{1-x}Si_x$ film. The details of the experiments are shown in Ref. 107.

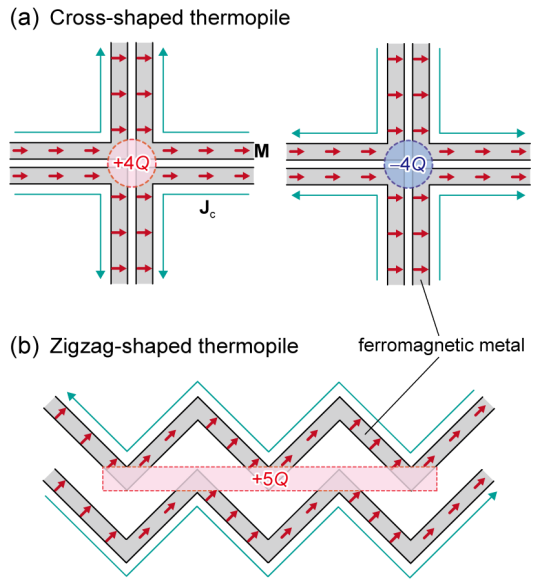

**Fig. 20.** Examples of the thermopile structures for AMPE: (a) cross-shaped thermopile and (b) zigzag-shaped thermopile. $Q$ denotes the AMPE-induced heat generated at the corner of a single wire. The total heat generated in the thermopile structures is proportional to the number of the integrated corners.

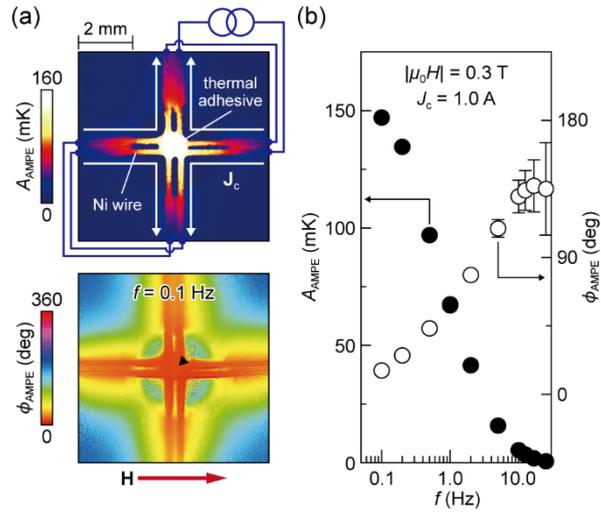

**Fig. 21.** (a) $A_{\rm AMPE}$ and $\phi_{\rm AMPE}$ images for the cross-shaped Ni thermopile at $|\mu_0 H| = 0.3$ T, $f = 0.1$ Hz, and $J_c = 1.0$ A. (b) $f$ dependence of $A_{\rm AMPE}$ and $\phi_{\rm AMPE}$ at the center of the cross-shaped Ni thermopile. The details of the experiments are shown in Ref. 108.

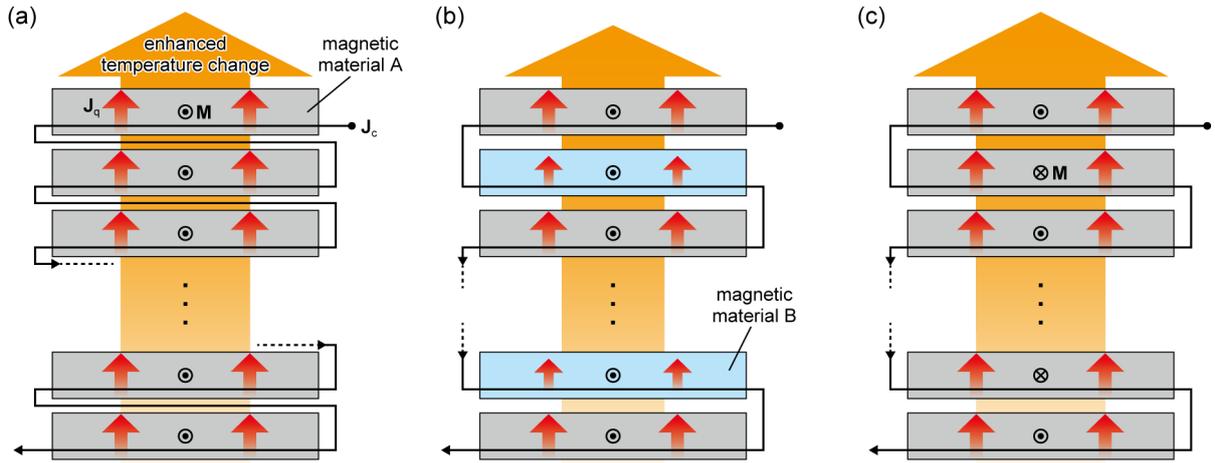

**Fig. 22.** Examples of the thermopile structures for AEE: (a) thermopile consisting only of a magnetic metal A in which **M** aligns along the same direction, (b) thermopile consisting of magnetic metals A and B with different anomalous Ettingshausen coefficients in which **M** aligns along the same direction, and (c) thermopile consisting only of a magnetic metal A in which **M** is reversed alternately.

Table I. Comparison between the LIT and LITR methods.

| | LIT | LITR |
|---|---|---|
| Spatial resolution (imaging) | ~ 10 μm | < 1 μm |
| Temperature resolution (typical values for 1h measurement) | 0.1 mK (imaging) | 0.1 mK (spot) <br> 1 mK (imaging) |
| Lock-in frequency | Min: no limit <br> Max: ~ 100 Hz | Min: no limit <br> Max: > 1 MHz |
| Temperature range | Min: ~ 270 K <br> Max: > 1000 K | Min: ~ 30 K <br> Max: > 1000 K |